\documentclass[12pt,a4paper,oneside]{extarticle}

\usepackage{cmap}
\usepackage[T2A]{fontenc}
\usepackage[utf8]{inputenc}
\usepackage[english]{babel}

\usepackage{multirow}

\usepackage{amsfonts}
\usepackage{amstext}
\usepackage{amsmath}
\usepackage{amssymb}
\usepackage{mathtools}
\usepackage{indentfirst} 

\usepackage[numbers,square,sort]{natbib} 
\bibpunct{(}{)}{,}{a}{,}{,} 

\usepackage{graphics}
\graphicspath{{graphics/}{pictures/}}
\usepackage[pdftex]{graphicx}
\usepackage{color}
\usepackage{lscape} 
\usepackage{bm} 
\usepackage{tikz}
\usepackage{titlesec} 
\usepackage{float} 
\usepackage{longtable}
\usepackage{booktabs}
\usepackage[center]{subfigure}
\usepackage{captcont}
\usepackage{tabularx}
\usepackage{diagbox} 
\usepackage{rotating}
\usepackage{geometry}
\usepackage[super]{nth}
\usepackage{comment} 
\usepackage{textcomp}


\newtheorem{remark}{Remark}

\newcommand{\BE}{\begin{equation}}
\newcommand{\EE}{\end{equation}}
\newcommand{\BEQ}{\begin{eqnarray}}
\newcommand{\EEQ}{\end{eqnarray}}
\newcommand{\BEQA}{\begin{eqnarray*}}
\newcommand{\EEQA}{\end{eqnarray*}}
\newcommand{\BA}{\begin{array}}
\newcommand{\EA}{\end{array}}

\newcommand{\CP}{\stackrel{p}{\longrightarrow}}

\newtheorem{theorem}{Theorem}
\newtheorem{lemma}{Lemma}
\newtheorem{proposition}{Proposition}
\newtheorem{assumption}{Assumption}

\begin{document}
\baselineskip= 7mm

\title{Time-Transformed Test for Bubbles under Non-stationary Volatility\footnote{The authors are grateful to two anonymous referees, Yang Zu, Rob Taylor, participants of the Center for Econometrics and Business Analysis (CEBA, St. Petersburg State University) seminar series, and seminars at Hitotsubashi University and Kyoto University for helpful comments and suggestions. Research of A. Skrobotov  was supported by a grant from the Russian Science Foundation (RSF, Project No. 20-78-10113). Research of A. Tsarev was supported by the RANEPA state assignment research program. Research of E. Kurozumi was supported by JSPS KAKENHI Grant Number 19K01585.}}
\author{
Eiji Kurozumi$^{a}$, Anton Skrobotov$^{b,c}$ and Alexey Tsarev$^{b}$\\
{\small {$^{a}$ Hitotsubashi University}}\\
{\small {$^{b}$ Russian Presidential Academy of National Economy and Public Administration}}\\
{\small {$^{c}$ Saint Petersburg State University, Center for Econometrics and Business Analysis}}
}

\maketitle

\begin{abstract}

This paper is devoted to testing for bubbles under time-varying non-stationary volatility. Because the limiting distribution of  the seminal \citet{PWY2011} test depends on the variance function and usually requires a bootstrap implementation under heteroskedasticity, we construct the test based on a deformation of the time domain. The proposed test is asymptotically pivotal under the null hypothesis and its limiting distribution coincides with that of the standard test under homoskedasticity, so that the test does not require computationally extensive methods for inference. Appealing finite sample properties are demonstrated through Monte-Carlo simulations. An empirical application demonstrates that the upsurge behaviour of cryptocurrency time series in the middle of the sample is partially explained by the volatility change.

\medskip

\noindent {Keywords: Rational bubble; Explosive autoregression; Time-varying volatility;
Right-tailed unit root testing; Variance profile; Time-transformed data}

\medskip

\noindent {\bf JEL Codes: C12, C22} 

\end{abstract}


\section{Introduction}

In this paper, we consider testing for bubbles under heteroskedastic volatility. Since a seminal paper of \citet{PWY2011} testing for and dating for rational bubbles have been investigated in both the theoretical and empirical econometrics literature. Because time series characterized by an explosive behaviour may be modeled by a process with an autoregressive time-varying coefficient, \cite{PWY2011} and \citet{PSY2015a,PSY2015b} proposed constructing subsample ADF-type statistics that take the supremum over the subsamples. \cite{HB2012} compared the subsample ADF-type test with several other alternative tests, including  the Chow-type unit root statistics, while \cite{Whitehouse2019} considered the \cite{PWY2011} test with the GLS-type detrending. Note that these studies assumed  that volatility in the shocks is constant over the sample period. However, it is well known in the empirical finance literature that volatility in financial time series may not be constant but change depending on time. In this case, the above tests for bubbles may suffer from size distortion because the limiting distributions of the test statistics depend on the volatility structure, as demonstrated by \citet{Cavaliere2004a,CavaliereTaylor2007a,CavaliereTaylor2007b,CavaliereTaylor2008b}, \textit{inter alia}, in the unit root testing literature and by \citet{HLST2016} for the \citet{PWY2011} test.

To address this issue, several papers  proposed testing for the explosive behaviour under time-varying volatility. \citet{HLST2016} showed that the limiting distribution of the supremum ADF (SADF) test of \citet{PWY2011} depends on the so-called variance profile, which will be defined in Section 3, and proposed a wild bootstrap implementation to obtain the asymptotically correct size of the test. \citet{hafner2020testing} modified the wild bootstrap algorithm of \citet{HLST2016} to allow the skewed distribution of the series. \citet{HLZ2019} proposed a weighted least squares based modification of the SADF test with a non-parametric kernel smoothing estimator of the variance process (their test is called SBZ test). The limit distribution of the test still depends on the variance profile under the null and local alternatives and the test also has a non-monotonic power problem in some cases. \citet{HLZ2019} suggested to use a union of rejections testing strategy based on two tests, the SBZ and the SADF (under this strategy, we reject the null of a unit root if at least one of the tests in a union rejects the null), coupling with a wild bootstrap implementation of \citet{HLST2016}. \citet{harvey2020sign} proposed another method which controls size under time-varying volatility. This method is based on the cumulated signs of the first differences of the dependent variable and leads to the pivotal null limiting distribution under time-varying volatility. However, under the local alternative, the sign-based test depends on volatility function, and \citet{harvey2020sign} suggested to use a wild bootstrap based union of rejections testing strategy with the SADF and sign-based tests. However, these tests are computationally expensive, and computation time increases rapidly if the sample size increases. Finally, \citet{AHLTZ2021} developed a CUSUM-based monitoring procedure for explosive episodes which is robust to time-varying volatility.

In this paper, we propose a new method for testing the explosive  behaviour under nonstationary volatility ($\sigma_{t}$) by transforming the series according to the volatility behaviour, similarly to \citet{CavaliereTaylor2007b} in the unit root testing context. More precisely, we take the sampling interval longer in the low volatility regime, whereas we take it shorter for large values of $\sigma_t$. We construct the test statistic based on transformed data that has the same limiting distribution as its standard counterpart constructed under homoskedasticity. The transformation of the time series is based on the variance profile estimated from data. For this, we utilise the approach of \cite{HLZ2021}; we estimate non-parametrically the time-varying autoregressive coefficient, collect residuals, and use them for estimating the variance profile. The supremum based test for the bubbles is constructed in a simple way based on the transformed version of the subsample tests. Monte-Carlo simulations demonstrate that our test proposed in this paper can control the empirical size well and is most powerful among the other existing tests in some cases, although the latter is not always the case and our test is less powerful in other cases. Thus, the contribution of this paper is the proposal of new tests with well controlled size. As a by-product of using the asymptotic critical values, our method is computationally less expensive comparing to the existing methods with the bootstrap method. An empirical application demonstrates that it is important to take non-stationary volatility into account when testing for a bubble in the time-series possibly with unstable volatility.

The paper is organized as follows. Section 2 formulates the model and assumptions. In Section 3 we propose to implement a deformation of the time domain for the time series and use the transformed series to construct the SADF type test statistic for the explosive behaviour. Using this time deformation with the variance profile assumed to be known, we show that the limiting distribution of the test statistic is the same as in the case of homoskedasticity. Section 4 discusses the case of the unknown variance profile and the corresponding limiting result. Monte-Carlo simulations are performed in Section 5. Section 6 gives the empirical application, and Section 7 concludes the paper.

\section{Model and Assumptions}

Consider the time series $\{y_t\}$ generated by the following data generating process (DGP) that allows one explosive regime with a subsequent collapsing regime: 
\begin{eqnarray}
  y_t &=& \mu+u_t,  \label{Explosive1}\\
  u_t &=& 
\begin{cases}  
u_{t-1}+\varepsilon_t, & t=1,\dots,\lfloor \tau_{1,0}T\rfloor,\\
(1+\delta_{1})u_{t-1}+\varepsilon_t, & t=\lfloor \tau_{1,0}T\rfloor+1,\dots,\lfloor \tau_{2,0}T\rfloor,\\
(1-\delta_{2})u_{t-1}+\varepsilon_t, & t=\lfloor \tau_{2,0}T\rfloor+1,\dots,\lfloor \tau_{3,0}T\rfloor,\\
u_{t-1}+\varepsilon_t, & t=\lfloor \tau_{3,0}T\rfloor+1,\dots,T,\\
\end{cases} \label{Explosive-u} \\
\varepsilon_t&=&\sigma_te_t
\label{Explosive2}
\end{eqnarray}
where $\delta_{1}\geq 0$, $\delta_{2}\geq 0$, $0 \leq \tau_{1,0}<\tau_{2,0}\leq \tau_{3,0}\leq 1$, and the initial value condition is given by $u_0=o_p(\sqrt{T})$. We assume that $\mu=0$ for simplicity but a positive shrinking drift may be allowed as considered by \citet{PSY2014} and \citet{PSY2015a,PSY2015b}. The process $\{y_t\}$ evolves as a unit root process, but a bubble possibly emerges at $\lfloor \tau_{1,0}T\rfloor+1$ ($\lfloor\cdot\rfloor$ denotes an integer part of value) with the explosive AR(1) coefficient given by $1+\delta_{1}$, followed by the collapsing regime from $\lfloor \tau_{2,0}T\rfloor+1$ to $\lfloor \tau_{3,0}T\rfloor$ generated as a stationary process, which is interpreted as the return to the normal market behaviour. The magnitude of $\delta_{2}$ specifies the extent of the collapse of the bubble with the duration between $\lfloor \tau_{2,0}T\rfloor + 1$ and $\lfloor \tau_{3,0}T\rfloor$. When a crash occurs, we assume, following \citet{HLZ2021}, that $y_{\lfloor \tau_{3,0}T\rfloor}\coloneqq y^*a_T$, where $y^*=O_p(1)$ and  $a_T=1$ or $a_T\to \infty$. The volatility of the innovations is given by $\sigma_t$ and it can be non-stationary. Note that we can simply rewrite the process for $\{y_t\}$ as
\begin{equation}\label{rho-t}
    y_t=(1+\delta_t)y_{t-1}+\varepsilon_t\qquad\mbox{or}\qquad \Delta y_t=\delta_t y_{t-1}+\varepsilon_t
\end{equation}
with obvious definition of $\delta_t$. The null hypothesis we consider is that the market is efficient and thus $\delta_{t}=0$ in expression \eqref{rho-t}.\footnote{The null hypothesis can be expressed using \eqref{Explosive-u} in several ways such that $\tau_{1,0}=1$, $\delta_{1}=0$ and $\tau_{2,0}=1$, or $\delta_{1}=\delta_2=0$.} On the other hand, we assume the existence of a bubble under the alternative, which corresponds to the case where $\delta_t$ in \eqref{rho-t} is not stable at 1 and the model is given by \eqref{Explosive1}--\eqref{Explosive2} with $\delta_{1}>0$.

We suppose that the following assumption holds.
\begin{assumption}\label{assumption-1}
(a) $\{e_t\}$ is a martingale difference sequence with respect to $\mathcal{F}_t\coloneqq \sigma(e_t,e_{t-1},\ldots)$ with $E[e_t^2|\mathcal{F}_{t-1}]=1$ and $E[|e_t|^p|\mathcal{F}_{t-1}] < \infty$ almost surely for some $p>6$. (b) The volatility $\sigma_t$ is defined as $\sigma_t\coloneqq \omega(t/T)$, where $\omega(s)\in D[0,1]$ for $s\in[0,1]$ is a non-stochastic and strictly positive function satisfying $0 < \underline{\omega} < \omega(s)<\bar{\omega} < \infty$.
\end{assumption}

Assumption \ref{assumption-1}(a) requires the existence of more than 6-th order moments, which is used for the nonparametric estimation of the AR(1) coefficient. Assumption \ref{assumption-1}(b) allows a general class of volatility processes; we can allow breaks in volatility, trending volatility, and regime switching volatility. See, for example, \citet{CavaliereTaylor2007a,CavaliereTaylor2007b}. 
Note that under homoskedasticity, $\sigma_t$ reduces to $\sigma$ for all $t$. Under Assumption \ref{assumption-1}, the conditional variance of $\varepsilon_t$ is given by $E[\varepsilon_t|{\cal F}_{t-1}]=\sigma_t^2$ and thus conditional heteroskedasticity is allowed in the model. Although the conditional and unconditional variances are the same ($\sigma_t^2$) under Assumption \ref{assumption-1}, our model can flexibly capture the typical property of financial data such as volatility clustering because $\sigma_t$ covers a wide class of nonlinear functions.

Under Assumption \ref{assumption-1}, the partial sum process of $\{\varepsilon_t\}$ is asymptotically characterized by the variance profile, termed by \citet{CavaliereTaylor2007a}, which is defined as
\[\eta(s)\coloneqq\left(\int_0^1\omega(r)^2dr\right)^{-1}\int_0^s\omega(r)^2dr.\]
We also define the so called (asymptotic) average innovation variance as
\[\bar{\omega}^2\coloneqq\int_0^1\omega(r)^2dr.\]
Note that $\eta(s)=s$ under homoskedasticity. Then, under Assumption \ref{assumption-1}, we have the following weak convergence due to Theorem 1 of \citet{CavaliereTaylor2007a}:
\begin{equation}
\label{fclt1}
\frac{1}{\sqrt{T}}y_{\lfloor rT\rfloor} =\frac{1}{\sqrt{T}}u_{\lfloor rT\rfloor} =\frac{1}{\sqrt{T}}\sum_{t=1}^{\lfloor rT\rfloor}\varepsilon_t\Rightarrow \bar{\omega}W(\eta(r))\eqqcolon\bar{\omega}W^\eta(r)\qquad(0\leq r \leq 1),
\end{equation}
where $\Rightarrow$ denote weak convergence in $D[0,1]$ and $W(\cdot)$ is a standard Brownian motion, while $W^\eta(\cdot)$ is called a variance transformed Brownian motion (Brownian motion under a modification of the time domain)\footnote{We thank one of the referees who pointed out that this Brownian motion can be expressed as Dambis-Dubins-Schwarz Brownian motion. See, for example, \cite{KS1991}.}. Again, we can see that in the case of a constant variance with $\sigma_t=\sigma$, we have $\eta(r)=r$ and thus $W^{\eta}(r)$ reduces to a standard Brownian motion.

\section{Deformation of Time Domain When the Variance Profile is Known}

In this section, we propose new tests for a bubble under the infeasible assumption that the variance profile is known. We will show that the new test statistics are asymptotically pivotal and thus can be implemented with the asymptotic critical values. The feasible versions will be considered in the next section.

To propose new test statistics, we first review the methodology proposed by \citet{PWY2011}, which utilize the maximum of the ADF test statistics constructed using subsamples. Let $ADF_{r_1}^{r_2}$ be the ADF $t$-statistic of $\delta$ in the regression
\begin{equation}\label{Cript1}
    \Delta y_t=\mu+\delta y_{t-1}+\varepsilon_t
\end{equation}
for $t=\lfloor r_1T\rfloor+1$ to $\lfloor r_2T\rfloor$. Then, the so-called SADF test statistic is defined as
\begin{equation}\label{TT2}
    SADF\coloneqq\sup_{r_2\in[r_0,1]}ADF_{0}^{r_2}
\end{equation}
where the right-tail is the rejection region. Furthermore, the SADF test is extended to multiple bubble context in \citet{PSY2015a,PSY2015b} and the generalized SADF (GSADF) test statistic is given by
\begin{equation}\label{TT3}
    GSADF(r_0)\coloneqq\sup_{r_2\in[r_0,1],r_1\in[0,r_2-r_0]}ADF_{r_1}^{r_2}.
\end{equation}
In this form, $r_{\omega}\coloneqq r_2-r_1$ is a window size with the minimum size as $r_0$. That is, for every given $r_2$, the ADF test statistic is calculated over all possible $r_1$ from 0 to $r_2-r_0$ and the GSADF test is constructed as the maximum of them. Note that the SADF test is the special case of the GSADF test by setting $r_1=0$ and $r_2=r_{\omega}\in[r_0,1]$.

These two tests are based on regression \eqref{Cript1} and we typically include a constant even if we assume $\mu=0$, because the effect of the initial value disappears under the null hypothesis by including a constant. We call this method the OLS-demeaning. On the other hand, \citet{Whitehouse2019} proposed the GLS-type demeaning, which  results in the increase of power in some cases.

Although the SADF and GSADF test statistics are asymptotically pivotal under the assumption of homoskedasticity, it is not difficult to see, as demonstrated by \citet{HLST2016} and \citet{HLZ2019}, that their limiting distributions depend on the volatility process through the variance profile $\eta(s)$ under the existence of  heteroskedasticity. In the following, we briefly investigate this dependence to explain our motivation for the construction of new test statistics.

Following \citet{HLZ2019}, we consider the SADF test statistic with a version of the GLS-type demeaning; we subtract the initial value of $y$ from all the observations, which is given by $\check{y}_t\coloneqq y_t-y_0$. We call this specific treatment the GLS-demeaning. In this case, the test statistic is constructed from \eqref{Cript1} without a constant and thus  $ADF_{r_1}^{r_2}$can be written as
\begin{equation}\label{DFr1r2}
ADF_{r_1}^{r_2}\coloneqq\frac{\sum_{t=\lfloor r_1T\rfloor+1}^{\lfloor r_2T\rfloor}\check{y}_{t-1}\Delta \check{y}_t}{\sqrt{\hat{\sigma}^2( r_1, r_2)\sum_{t=\lfloor r_1T\rfloor+1}^{\lfloor r_2T\rfloor}\check{y}_{t-1}^{2}}},
\end{equation}
where $\hat{\sigma}^2( r_1, r_2)$ is the usual variance estimator based on the regression residuals in this subsample. From \eqref{fclt1}, we can deduce that under the null hypothesis,
\begin{eqnarray}
\label{df:num}
\frac{1}{T}\sum_{t=\lfloor r_1T\rfloor+1}^{\lfloor r_2T\rfloor}\check{y}_{t-1}\Delta \check{y}_t&=&\frac{1}{2T}\left[\check{y}_{\lfloor  r_2T\rfloor}^{2}-\check{y}_{\lfloor  r_1T\rfloor}^{2}-\sum_{t=\lfloor r_1T\rfloor+1}^{\lfloor r_2T\rfloor}(\check{y}_t-\check{y}_{t-1})^2\right]\\
&\Rightarrow&  \frac{1}{2}\left(\bar{\omega}^2W^\eta( r_2)^2-\bar{\omega}^2W^\eta( r_1)^2-\int_{ r_1}^{ r_2}\omega(r)^2dr\right),
\label{df:num2} \\
\frac{1}{T^2}\sum_{t=\lfloor r_1T\rfloor+1}^{\lfloor r_2T\rfloor}\check{y}_{t-1}^{2}&\Rightarrow& \bar{\omega}^2\int_{ r_1}^{ r_2}(W^\eta(r))^2dr,
\label{df:den1}\\
\hat{\sigma}^2( r_1, r_2)&\rightarrow_p& \frac{1}{ r_2- r_1}\int_{ r_1}^{ r_2}\omega(r)^2dr.
\label{df:den2}
\end{eqnarray}
As a result, we obtain the following proposition, which is also stated in \citet{HLZ2019}.

\begin{proposition}
Suppose that  Assumption 1 holds. Under the null hypothesis, we have, with the GLS-demeaning,
\[
SADF\Rightarrow \sup_{r_2\in[r_0,1]}\widetilde{ADF}_{0}^{r_2}(\omega)\quad\mbox{and}\quad GSADF(r_0)\Rightarrow\sup_{r_2\in[r_0,1],r_1\in[0,r_2-r_0]}\widetilde{ADF}_{r_1}^{r_2}(\omega),
\]
\begin{equation}
\label{df:limit1}
\mbox{where}\quad \widetilde{ADF}_{r_1}^{r_2}(\omega)\coloneqq \frac{\bar{\omega}\left(W^\eta( r_2)^2-W^\eta( r_1)^2-\cfrac{\int_{ r_1}^{ r_2}\omega(r)^2dr}{\bar{\omega}^2}\right)}{2\sqrt{\frac{1}{ r_2- r_1}\int_{ r_1}^{ r_2}\omega(r)^2dr}\sqrt{\int_{ r_1}^{ r_2}(W^\eta(r))^2dr}}.
\end{equation}
\end{proposition}

We can see that the limiting distribution of the $ADF_{r_1}^{r_2}$ statistic depends on a variance-transformed Brownian motion, which in turn depends on the variance profile. Similarly, $ADF_{r_1}^{r_2}$ with the OLS-demeaning also depends on the variance structure as shown by \citet{HLST2016}. As a result, we cannot control the size of the test by using the critical values obtained under the assumption of homoskedasticity. Note that, in the special case of homoskedasticity with $\sigma_t=\sigma$, $\widetilde{ADF}_{r_1}^{r_2}$ reduces to the well-known form given by
\begin{equation}
\label{df:limit0}
\widetilde{ADF}_{r_1}^{r_2}(1)\coloneqq \frac{W( r_2)^2-W( r_1)^2-( r_2- r_1)}{2\sqrt{\int_{ r_1}^{ r_2}W(r)^2dr}},
\end{equation}
and thus, the distribution becomes free of the variance profile in this case.

To overcome this problem, \citet{HLST2016} and \cite{HLZ2019} proposed to use the wild bootstrap method and showed that the size of the test can be controlled. However, this method can be time-consuming if the sample size is very large. Instead, we propose the alternative method based on the time transformation of the series introduced by \citet{CavaliereTaylor2007b}. In this method, we take the sampling interval longer in the low volatility regime, whereas we take it shorter for large values of $\sigma_t$. As a result, the variation in the increment of each interval becomes stable as if the series were generated from the innovations with a constant variance and thus, we will have a standard result.


More precisely, the time transformation is based on the variance profile $\eta(s)$. We first note that, because the variance profile is a strictly monotonically increasing function under Assumption \ref{assumption-1}, we have the unique inverse given by $g(s)\coloneqq\eta^{-1}(s)$. Then, consider the time-transformed series $\tilde{y}_t=y_{t'}-y_{t'=0}$ with a non-decreasing sequence $t'=\lfloor g(t/T)T\rfloor$. We note that $\tilde{y}_0=y_0-y_0=0$ and $\tilde{y}_T=y_T-y_0$.\footnote{We can also consider the same transformation with the OLS-demeaning. However, our preliminary simulations show that the tests with the OLS-demeaning are inferior to those with the GLS-demeaning in view of power. More precisely, the former tests suffer from non-monotonic power in the sense that the power does not necessarily increase as the size of the bubble gets larger. Therefore, we focus on only the tests with the GLS-demeaning.} As shown by (9) in \citet{CavaliereTaylor2007b}, we have, under the null hypothesis,
\begin{equation}
\label{fclt:2}
T^{-1/2}\tilde{y}_{\lfloor r T\rfloor}\approx T^{-1/2} y_{\lfloor g(\lfloor r T\rfloor/T)T\rfloor}\approx T^{-1/2}y_{\lfloor g(r)T\rfloor} \Rightarrow \bar{\omega}W^{\eta}(g(r))=\bar{\omega}W(r)
\end{equation}
because $W^{\eta}(g(r))=W(\eta(g(r)))=W(r)$, and thus the time-transformed series behaves as if it were a unit root process with a constant variance. 

By taking into account \eqref{fclt:2} and the fact that the numerator of $ADF_{r_1}^{r_2}$ can be expressed as \eqref{df:num}, we propose the following test statistics based on the time-transformed ADF statistics:
\[
   STADF\coloneqq\sup_{r_2\in[r_0,1]}TADF_{0}^{r_2}
\qquad\mbox{and}\qquad
   GSTADF(r_0)\coloneqq\sup_{r_2\in[r_0,1],r_1\in[0,r_2-r_0]}TADF_{r_1}^{r_2},
\]
\begin{equation}\label{TT4}
\mbox{where}\quad
TADF_{r_1}^{r_2}=\frac{\tilde{y}_{\lfloor  r_2T\rfloor}^2-\tilde{y}_{\lfloor  r_1T\rfloor}^2-\bar{\omega}^2(\lfloor  r_2T\rfloor-\lfloor  r_1T\rfloor)}{2\bar{\omega}\sqrt{\sum_{t=\lfloor  r_1T\rfloor+1}^{\lfloor  r_2T\rfloor}\tilde{y}_{t-1}^2}}.    
\end{equation}
It is not difficult to derive the limiting distributions of the test statistics under the null hypothesis by using \eqref{fclt:2}. On the other hand, equation \eqref{df:num} is valid only under the null hypothesis and the behaviour of the test statistics is not obvious under the alternative. We summarize the asymptotic behaviour of the new tests in the following theorem.

\begin{theorem}\label{theorem-consistency}
Suppose that Assumption \ref{assumption-1} holds. (i) Under the null hypothesis,
\[
STADF\Rightarrow \sup_{r_2\in[r_0,1]}\widetilde{ADF}_{0}^{r_2}(1)\quad\mbox{and}\quad GSTADF(r_0)\Rightarrow\sup_{r_2\in[r_0,1],r_1\in[0,r_2-r_0]}\widetilde{ADF}_{r_1}^{r_2}(1).
\]
(ii) Under the alternative with $\delta_{1}>0$ and $\delta_{2}\geq 0$, both $STADF$ and $GSTADF$ are $O_p(\sqrt{T}\phi_1^{(\tau_{2,0}-\tau_{1,0})T})$.
\end{theorem}

From Theorem \ref{theorem-consistency}(i), we can see that the limiting distribution of $TADF_{r_1}^{r_2}$ under Assumption \ref{assumption-1} coincides with \eqref{df:limit0}, the limiting distribution of $ADF_{r_1}^{r_2}$ under assumption of homoskedasticity, and thus we obtain the pivotal limiting distributions of the test statistics. Therefore, we can use the same critical values as provided by \citet{Whitehouse2019} for the GLS demeaning case\footnote{For $r_0=0.1$, they are equal to 2.319, 2.626, 3.223 for 10\%, 5\% and 1\% significance levels. The critical values for other values $r_0$ can be easily computed from the R-code available in https://sites.google.com/site/antonskrobotov/}. We call these tests as STADF and GSTADF tests. Theorem \ref{theorem-consistency}(ii) implies that the test statistics diverge to infinity as $T\to \infty$ because $\tau_{2,0}-\tau_{1,0} > 0$ and the dominant term is positive, as shown in the proof. Because the rejection regions are right-hand side, this theorem implies that the tests are consistent.

\section{The Case of Unknown Variance Profile}

In practice, the variance profile $\eta(s)$ is generally unknown and we have to estimate it from the data. As \citet{CavaliereTaylor2007b} suggested, we  would like to construct $\hat{\eta}(s)$ from the estimator of $\varepsilon_t$, but we cannot use the regression residuals from of $\Delta y_t$ on $y_{t-1}$ because the true regression coefficient is time dependent under the alternative (the sample period includes explosive and collapsing regimes), which may result in the decrease of power. Instead, we utilize the approach taken by \citet{HLZ2021} and use the kernel-type local least squares method to estimate the time varying parameter $\delta_t$ in \eqref{rho-t} as follows:
\BEQ
\hat{\delta}_{t}
& = &
\arg\min_{\delta}\sum_{i=1}^TG_h(i/T-t)(\Delta \check{y}_i-\delta \check{y}_{i-1})^2 \nonumber \\
& = &
\left(\sum_{i=1}^TG_h(i/T-t)\check{y}_{i-1}^2\right)^{-1}\left(\sum_{i=1}^TG_h(i/T-t)\check{y}_{i-1}\Delta \check{y}_i\right),
\label{delta:lls}
\EEQ
where $G_h(s)=(1/h)G(s/h)$ and $\check{y}_i=y_i-y_0$. For this kernel $G(\cdot)$ and the bandwidth, we make the following assumption.

\begin{assumption}
\label{assumption:G}
 (a) $G(\cdot)$ is strictly positive only on $[-1,1]$ and $0 < \int_{-1}^1G(s)ds=\gamma < \infty$. \\
(b) $h\to 0$ as $T\to \infty$ with
\[
\frac{log(T)}{T^{1-4/(p-1)}h^{1-2/(p-1)}} \to 0.
\]
\end{assumption}

Next, define $\hat{\varepsilon}_{t}^*=\hat{\varepsilon}_t 1(|\hat{\varepsilon}_t| < \psi_T)$ as considered in \citet{HLZ2021} where $\hat{\varepsilon}_t=\Delta \check{y}_t-\hat{\delta}_t\check{y}_{t-1}$ and $\psi_T$ is a truncation parameter.  Then, we construct the estimator of the variance profile as considered by \citet{CavaliereTaylor2007b}:
\BE
\hat{\eta}(s)=\frac{\displaystyle \sum_{t=1}^{\lfloor sT \rfloor}\hat{\varepsilon}_t^{*2}+(sT-\lfloor sT \rfloor)\hat{\varepsilon}_{\lfloor sT \rfloor+1}^{*2}}{\displaystyle \sum_{t=1}^T\hat{\varepsilon}_t^{*2}}.
\label{VP:est}
\EE
The truncation parameter $\psi_T$ satisfies the following condition.
\begin{assumption}
\label{assumption:psi}
$\psi_T\ \to \infty$ with
\[
\frac{T}{\psi_T^p}\to 0\quad\mbox{and}\quad h\psi_T^2\to 0.
\]
\end{assumption}


By defining $\hat{g}(s)\coloneqq\hat{\eta}^{-1}(s)\coloneqq\inf\{u:\hat{\eta}(u)\geq s\}$, we have the following proposition.

\begin{proposition}\label{prop-consistency-estimator}
(i) Under Assumptions \ref{assumption-1}--\ref{assumption:psi}, $\hat{\eta}(s)-\eta(s)\CP 0$ uniformly over $s\in[0,1]$ under both the null and the alternative.  
(ii) $T^{-1/2}y_{\lfloor \hat{g}(s)T \rfloor}\Rightarrow \bar{\omega} W(s)$ under the null hypothesis.
\end{proposition}

Based on the result of Proposition \ref{prop-consistency-estimator}, we construct the test statistics $TADF_{r_1}^{r_2}$ as in \eqref{TT4} with $g(\cdot)$ and $\bar{\omega}$ replaced by the corresponding estimators. That is, allowing for an abuse of notation, we redefine $TADF_{r_1}^{r_2}$ as 
\[
TADF_{r_1}^{r_2}=\frac{\tilde{y}_{\lfloor  r_2T\rfloor}^2-\tilde{y}_{\lfloor  r_1T\rfloor}^2-\hat{\bar{\omega}}^2(\lfloor  r_2T\rfloor-\lfloor  r_1T\rfloor)}{2\hat{\bar{\omega}}\sqrt{\sum_{t=\lfloor  r_1T\rfloor+1}^{\lfloor  r_2T\rfloor}\tilde{y}_{t-1}^2}},
\]
and construct $STADF$ and $GSTADF$ as before, where
\[
\tilde{y}_t=y_{\lfloor \hat{g}(t/T)T \rfloor}-y_0
\quad\mbox{and}\quad
\hat{\bar{\omega}}^2=\frac{1}{T}\sum_{t=1}^T\hat{\varepsilon}_t^{*2}.
\]

\begin{theorem}\label{theorem-ghat}
Suppose that Assumptions \ref{assumption-1}-\ref{assumption:psi} hold. Then, under the null hypothesis, $STADF$ and $GSTADF(r_0)$ have the same limiting distributions as given in Theorem \ref{theorem-consistency}.
\end{theorem}

\begin{remark}
\label{remark:1}
Our analysis is based on the time-varying AR(1) model as in \eqref{rho-t}, which has been used in empirical analyses, but it is more flexible to allow for serial correlation in $\{\varepsilon_t\}$. If $\delta_t$ were known, we could replace $\hat{\bar{\omega}}^2$ with $\hat{\sigma}_{e^*}^2/(1-\hat{\beta}(1))^2$, as considered by \citet{CavaliereTaylor2007a,CavaliereTaylor2007b},  where $\hat{\beta}(1)\coloneqq \sum_{i=1}^k\hat{\beta}_i$ and $\hat{\sigma}_{e^*}^2$ are obtained by the regression
\BE
(\Delta y_t-\delta_ty_{t-1})=\sum_{i=1}^k\hat{\beta}_i(\Delta y_{t-i}-\delta_{t-i} y_{t-i-1})+\hat{e}_t^*
\label{serialcorr}
\EE
for $t=1,\ldots, T$, while $\eta_t$ can be estimated as \eqref{VP:est} with $\hat{\varepsilon}_t^*$ replaced by $\Delta y_t-\delta_t y_{t-1}$. In practice, we need to replace $\delta_t$ with its estimator. One of the possible candidates may be $\delta_t=1$, which is the true value of $\delta_t$ under the null hypothesis. The other possibility is to use the local least squares estimator of $\delta_t$ as considered in \eqref{delta:lls} and to replace $\Delta y_t-\delta_ty_{t-1}$ with $\hat{\varepsilon}_t^*$ in regression \eqref{serialcorr} and $\eta_t$. 
\end{remark}

To implement the feasible versions of the tests in practice, we need to choose $h$ and $\psi_T$ that satisfy Assumptions \ref{assumption:G}(b) and  \ref{assumption:psi}. For example, when $E|\varepsilon|_t^8 <\infty$ ($p=8$), we can choose $h=T^{-2/5}$ and $\psi_T=T^{1/7}$, while for a kernel $G(\cdot)$, we may use the uniform kernel $G(u)=1(-1\leq u \leq 1)$ or the truncated Gaussian kernel $G(u)=(2\pi)^{-1/2}\exp(-u^2/2)1(-1\leq u \leq 1)$. The choice of the tuning parameters will be further discussed in the next section.

\section{Monte Carlo Simulations}

Now we investigate the finite sample size and power properties of the proposed $STADF$ test in comparison to the standard $SADF$ test, the wild bootstrap $SADF$ test of \citet{HLST2016} ($SADF_b$), a union of rejections of the $SADF_b$ and SBZ tests of \citet{HLZ2019} ($SBZ$), sign-based tests of \citet{harvey2020sign} ($sSADF$), and a union of rejections of the $SADF_b$ and $sSADF$ ($sSADF_u$). All experiments were based on 1,000 Monte Carlo replications. For the wild bootstrap tests, $B = 199$ bootstrap replications were used. For the standard $SADF$ test and the time-transformed test, $STADF$, the conventional asymptotic critical values (based on the assumption of homoskedasticity) were used.\footnote{All simulations were programmed in R, and code is available on https://sites.google.com/site/antonskrobotov/}

Monte-Carlo simulations reported in this section are based on data generated by \eqref{Explosive1}-\eqref{Explosive2} with $\mu=0$, $u_0=e_0$, $e_t\sim IIDN(0,1)$. Data were generated from this DGP for samples of $T=100$ and 200 with the volatility process $\sigma_t$ satisfying one of the following models:
\begin{enumerate}
    \item Single shift in volatility: $\omega(s) = \sigma_{0} + (\sigma_{1} - \sigma_{0})1(s > \tau_{\sigma})$, where $\tau_{\sigma} \in \{0.3, 0.5, 0.7\}$ and $\sigma_1/\sigma_0\in\{1/6,1/3,1,3,6\}$. Note that $\sigma_1/\sigma_0=1$ corresponds to the case of constant unconditional volatility.
    \item Double shift in volatility: $\omega(s) = \sigma_{0} + (\sigma_{1} - \sigma_{0})1(0.4 < s \leq 0.6)$. 
    \item Logistic smooth transition in volatility: $\omega(s) = \sigma_{0} + (\sigma_{1} - \sigma_{0})\frac{1}{1+exp\{-50(s-0.5)\}}$.
    \item Trending volatility: $\omega(s) = \sigma_{0} + (\sigma_{1} - \sigma_{0})s$. 
\end{enumerate}

To investigate the finite sample size and power, we consider the set of the bubble magnitude $\delta_1\in\{0,0.02,0.04,0.06,0.08,0.1\}$ and $\delta_2=0$ for the non-collapsing case, with the bubble occurring between $\lfloor\tau_{1,0}T\rfloor=0.4T$ and $\lfloor\tau_{1,0}T\rfloor=0.6T$.

For all tests, the parameters $r_0 = 0.01 + 1.8/\sqrt{T}$, $k=0$ (the number of lagged differences) and $seed = [10^4*\hat{\sigma}_{y}]$ (where $\hat{\sigma}_{y}$ is the standard deviation of $y_t$). We use the Gaussian kernel in variance estimation for the $SBZ$ and the uniform kernel in regression \eqref{delta:lls} for the $STADF$ test. For the $STADF$ test, the bandwidth $h$ in regression \eqref{delta:lls} is selected based on the leave-one-out cross-validation for $h$ being between $T^{-0.5}$ and $T^{-0.3}$, and the truncation parameter $\psi_T = \bar{\sigma}*T^{\frac{1}{7}}$, where $\bar{\sigma} = \max\limits_{s=1,2,\ldots,0.9T} \hat{\sigma}_{s}$ with $\hat{\sigma}_{s}$ being the standard deviation of the residuals from the local linear regressions, $\hat{\varepsilon}_t$, calculated in a subsample with $t=s,\ldots,(s+0.1T)$.\footnote{ This approach to selecting the parameter $\bar{\sigma}$ is motivated by but differs from \cite{HLZ2021}, in which $\bar{\sigma} = \hat{\sigma}_{\Delta y_{t}}$ for $t = 1, \ldots, 0.1T$ is used. Our preliminary simulations show that too small $\psi_T$ results in over-size distortion, whereas the tests become conservative for large $\psi_T$. We tried various versions of $\bar{\sigma}$ and found that the current formulation results in stable empirical size. Because we calculate $\bar{\sigma}$ based on the untruncated local regression residuals in various subsamples, $\bar{\sigma}$ tends to be greater than the corresponding value in \cite{HLZ2021} and thus gives a larger $\psi_T$ with less truncation.} This choice of the parameters ensures that Assumptions \ref{assumption:G}(b) and  \ref{assumption:psi} are fulfilled.


Tables~\ref{tab_1}--\ref{tab_3} present the Monte Carlo results for single volatility shift with $\tau_{\sigma}=0.5$, $\tau_{\sigma}=0.3$, and $\tau_{\sigma}=0.7$ respectively, and Tables~\ref{tab_4}--\ref{tab_6} for double shift in volatility, logistic smooth transition in volatility, and trending volatility, respectively. The numbers in these tables are the rejection frequencies of the tests at 5\% significance level. First, it should be noted that the power of each of the tests increases monotonically with $\delta_1$ for all sets of parameters $\sigma_0$ and $\sigma_1$. Second, the power of all robust tests increases with the number of observations $T$, which confirms their consistency. Third, the original SADF test suffers from severe size distortion under nonstationary volatility, as is observed in the existing literature.

In the single volatility shift case with $\tau_{\sigma}=0.5$ (Table~\ref{tab_1}) when $T=100$ and volatility decreases moderately with $\sigma_{1}/\sigma_{0} =1/3$, our test controls the empirical size well and is most powerful among others, whereas when $\sigma_{1}/\sigma_{0} =1/6$, the $sSADF$, $sSADF_u$ and $STADF$ tests are close to each other, and the differences depend on the extent of the explosive regime. 
The $SBZ$ test looks to have the second best power when $\sigma_{1}/\sigma_{0} =1/3$ but it suffers from oversize distortion.  For the homoscedastic case with $\sigma_{1}/\sigma_{0}=1$, the best test in terms of power is the $SBZ$ test, but the difference of power is relatively minor among the tests except $sSADF$. 
When volatility increases, i.e. $\sigma_{1}/\sigma_{0} > 1$, the $STADF$ test still control the size well but tends to be slightly less powerful than the others. The similar tendency is observed when $T=200$ but the $sSADF$ becomes as good as our test for $\sigma_{1}/\sigma_{0} =1/3$.

When the shift occurs at $\tau_{\sigma}=0.3$, the disadvantage of our test in view of power disappears and it becomes most powerful when $T=200$ with $\sigma_{1}/\sigma_{0} =6$ as is shown in Table~\ref{tab_2}, whereas it tends to be less powerful than the $sSADF$ test (and $sSADF_u$) when $\sigma_{1}/\sigma_{0} =1/6$. Similarly, as in Table~\ref{tab_3}, our test performs best when $\tau=0.7$ with $\sigma_{1}/\sigma_{0} =3$, while it is not the case with $\sigma_{1}/\sigma_{0} =6$.

Similar conclusions can be drawn if we analyze double volatility shift (see Table~\ref{tab_4}). More generally, when the volatility has a jump(s) as in the cases of single and double volatility shift, $STADF$ is more powerful (less powerful) than $sSADF$ if $\sigma_0 \leq \sigma_1$ ($\sigma_0 > \sigma_1$).

When the volatility changes smoothly as in the cases of  logistic smooth transitionin volatility (Table~\ref{tab_5}) and trending volatility (Table~\ref{tab_6}), our test can control the size relatively well in many cases and is most powerful among others in some cases, but the latter is not always the case; no test dominates the others in view of power throughout the simulations. As a whole, the standard $SADF$ test has an incorrect size in the case of nonstationary volatility. 
It is also worth noting that the power of the $STADF$ test is often higher than the power of the other robust tests for the small bubble magnitudes.
Closely comparing two heteroskedasticity robust asymptotic tests, $STADF$ and $sSADF$, in Tables~\ref{tab_5} and \ref{tab_6}, the $STADF$ is more powerful than $sSADF$. That is, in the case of the smooth change in volatility, our test has an advantage over $sSADF$ in most cases,
though this result seems to be partially due to the conservativeness of $sSADF$ in some cases.

Summarising the results, the $STADF$ has the better size and best power in some cases but is less powerful in other cases, although the differences are not necessarily large. Moreover, our test is not computationally expensive in comparison to the bootstrap-based tests. The importance of the latter grows if the sample size increases, because the bootstrap-based tests are quite time consuming to implement for very long time series with reasonable $B$. 

\begin{figure}[h]
    \centerline{\includegraphics[width=0.99\textwidth]{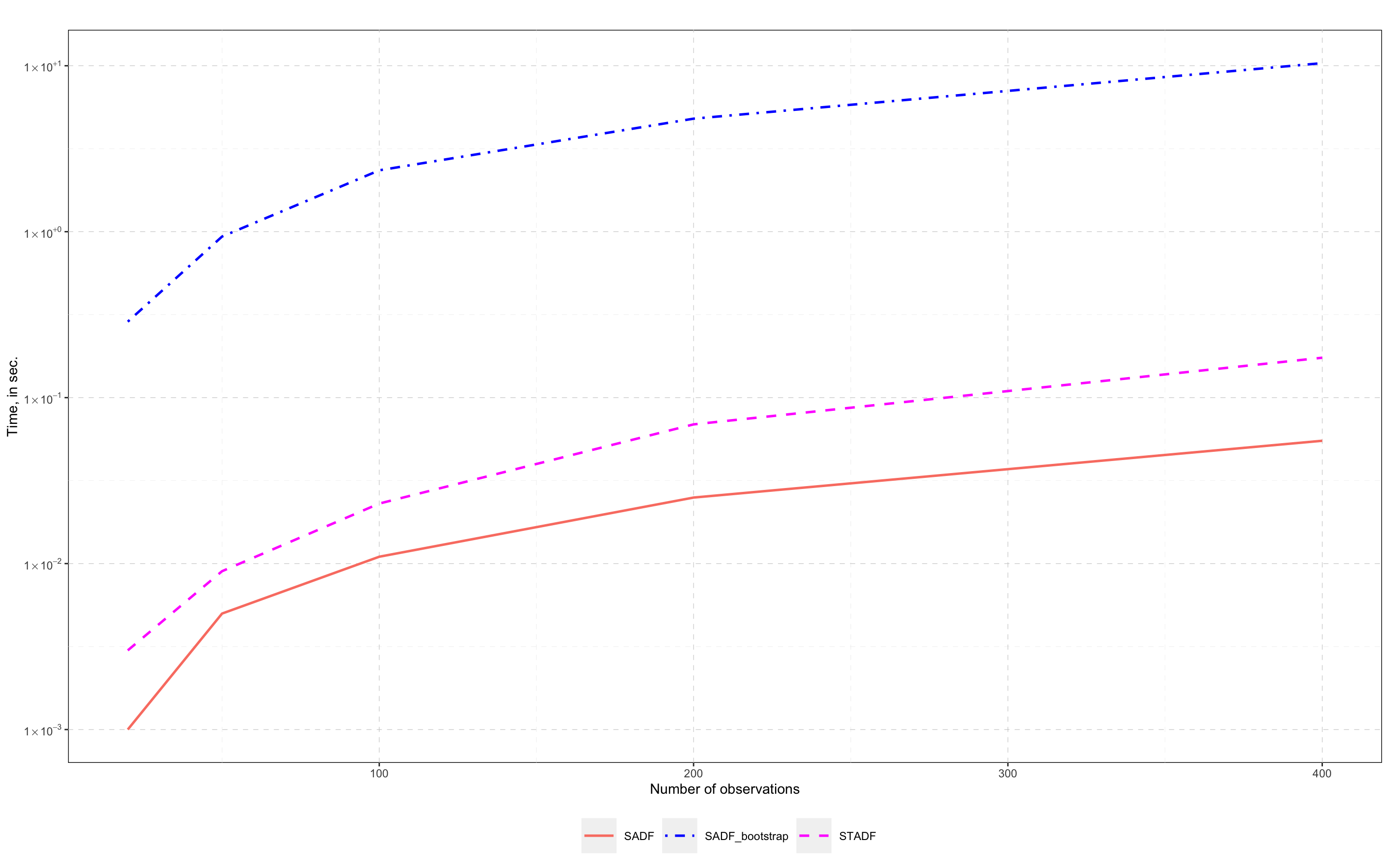}}
    \caption{Computational time of $SADF$, $SADF_{b}$ and $STADF$ tests} \label{time_graph}
\end{figure}

Figure~\ref{time_graph} demonstrates the comparative computation time of the $SADF$, $SADF_{b}$, and $STADF$ tests. For the simulation, we used a simple random walk process without bubbles or volatility shifts. We estimated the median computation time for 1000 iterations and for 20, 50, 100, 200 and 400 observations.\footnote{On a MacBook Pro laptop with a 3.1 GHz processor using R.} For $SADF_{b}$, we used $B = 199$. It should be noted that the computation times of the $STADF$ and $SADF_{b}$ tests differ dozens of times. For example, for $T = 400$, the median computation time of the $STADF$ test was 0.17 seconds, and the running time of the $SADF_{b}$ was 10.37 seconds, a difference of more than 59 times. At the same time, $STADF$ test is slightly slower than the $SADF$ test, which is to be expected. 

\section{Empirical application}

\begin{figure}[h]%
\begin{center}%
\subfigure[btc]{\includegraphics[width=0.31\linewidth]{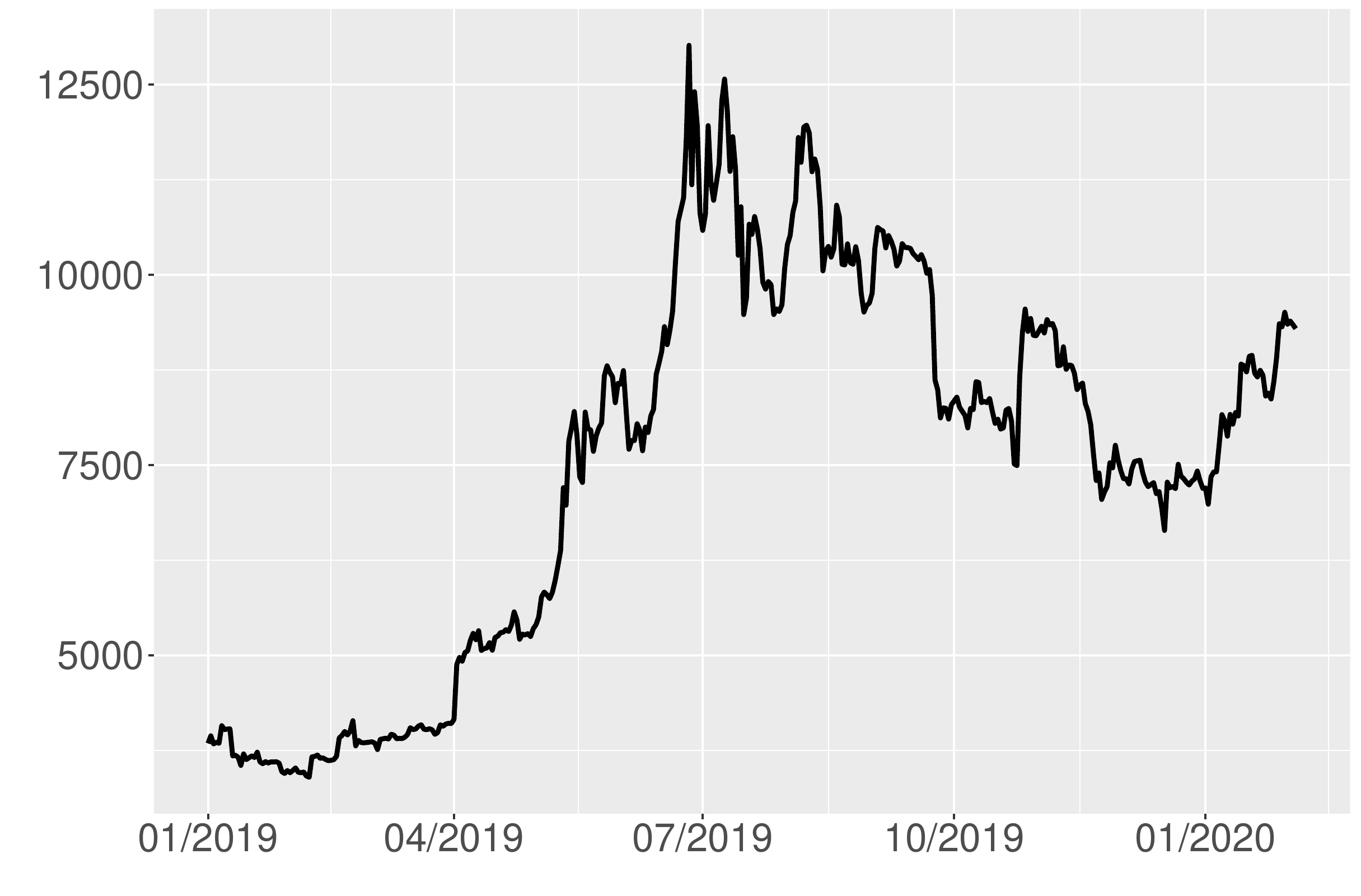}}
\subfigure[eth]{\includegraphics[width=0.31\linewidth]{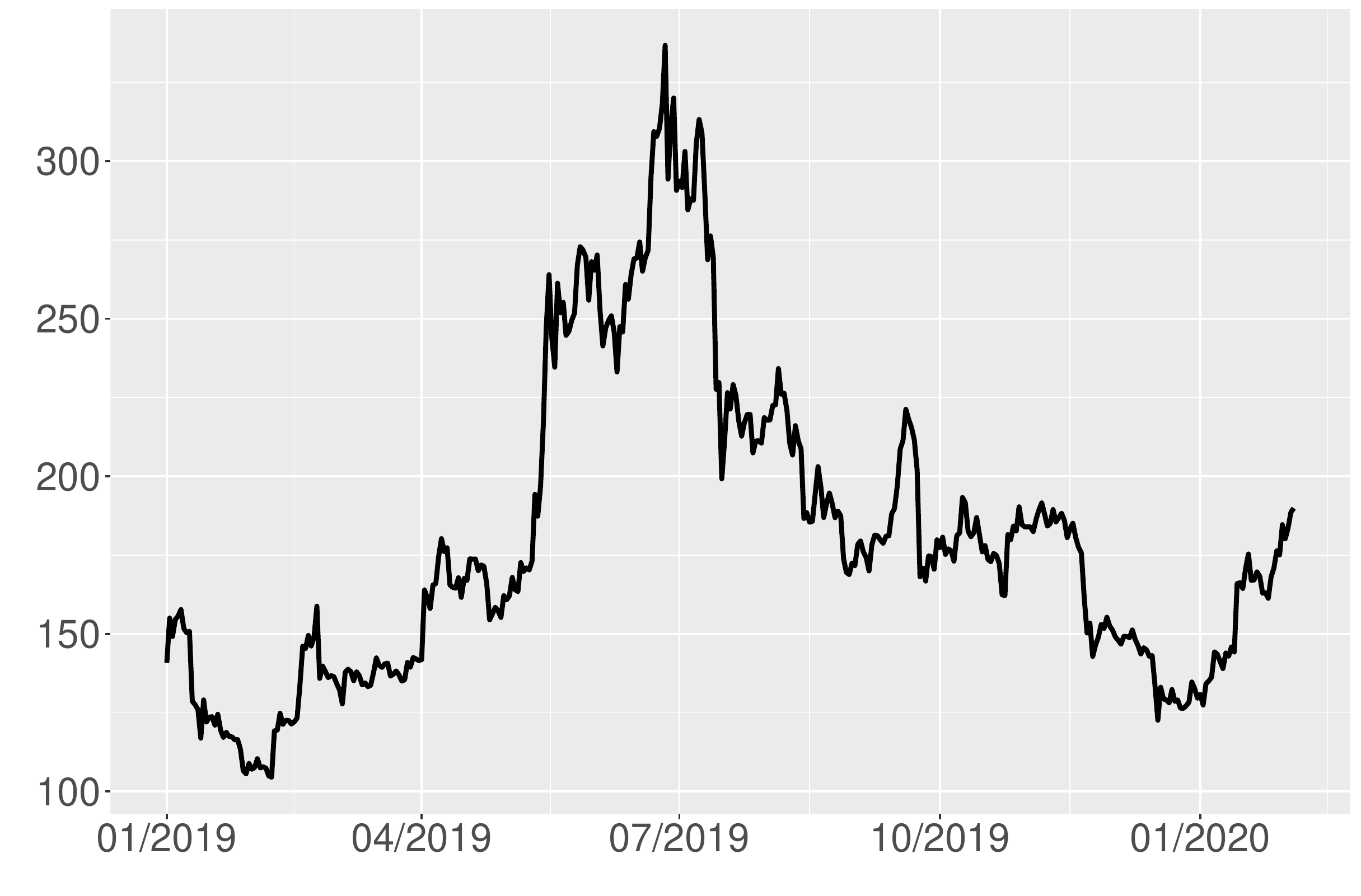}} 
\subfigure[xrp]{\includegraphics[width=0.31\linewidth]{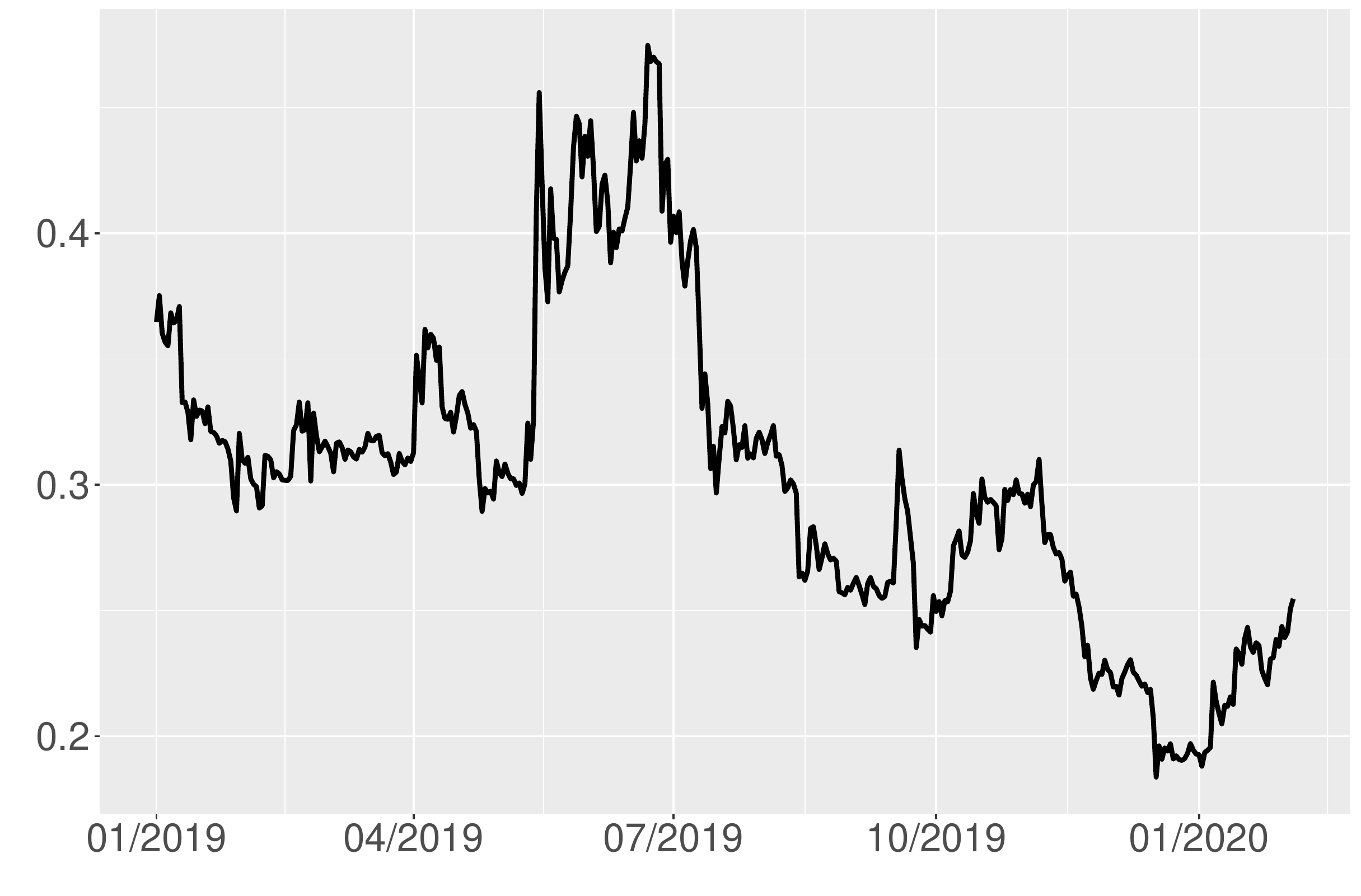}}
\subfigure[$\eta_{btc}$]{\includegraphics[width=0.31\linewidth]{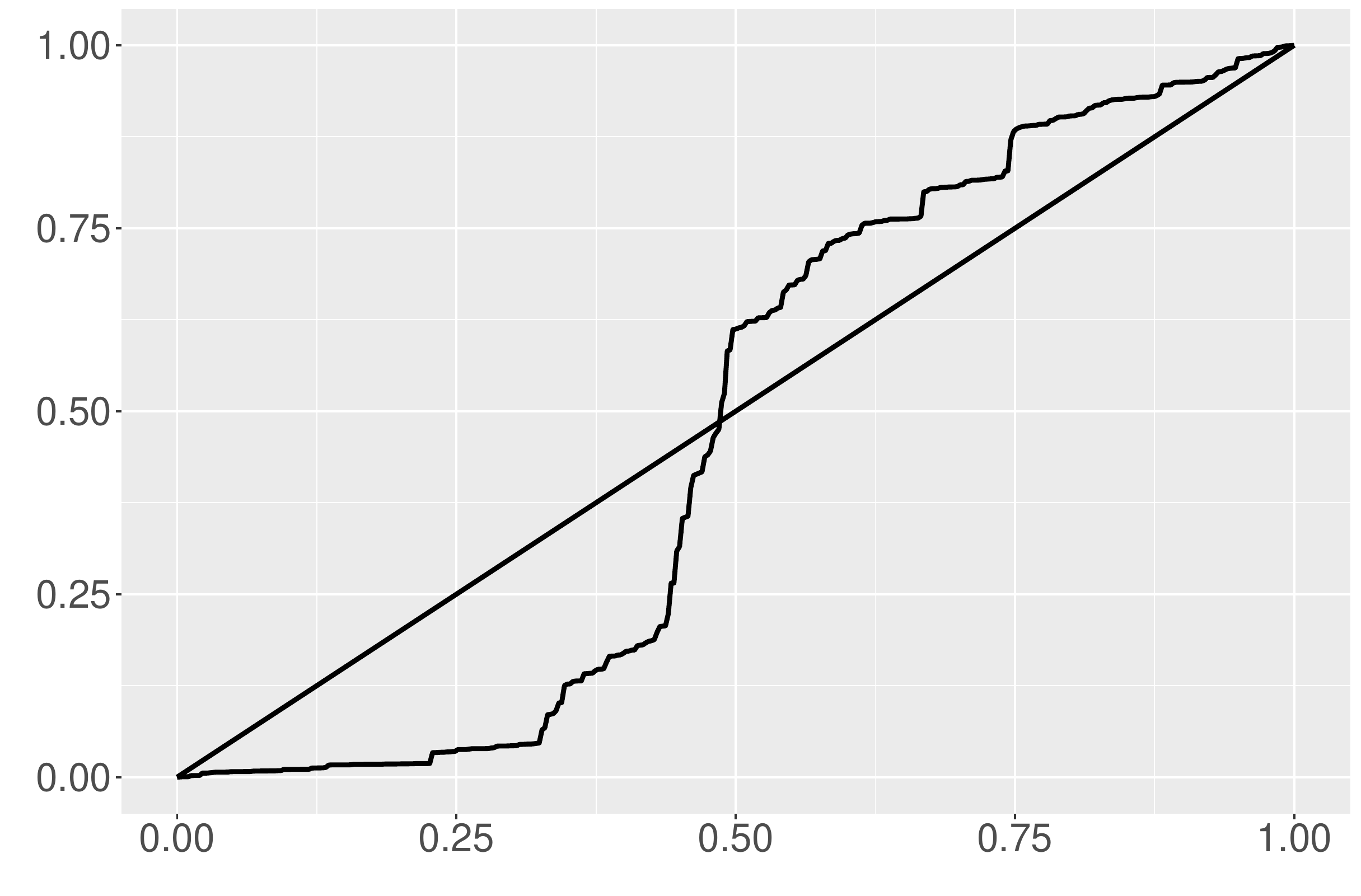}}
\subfigure[$\eta_{eth}$]{\includegraphics[width=0.31\linewidth]{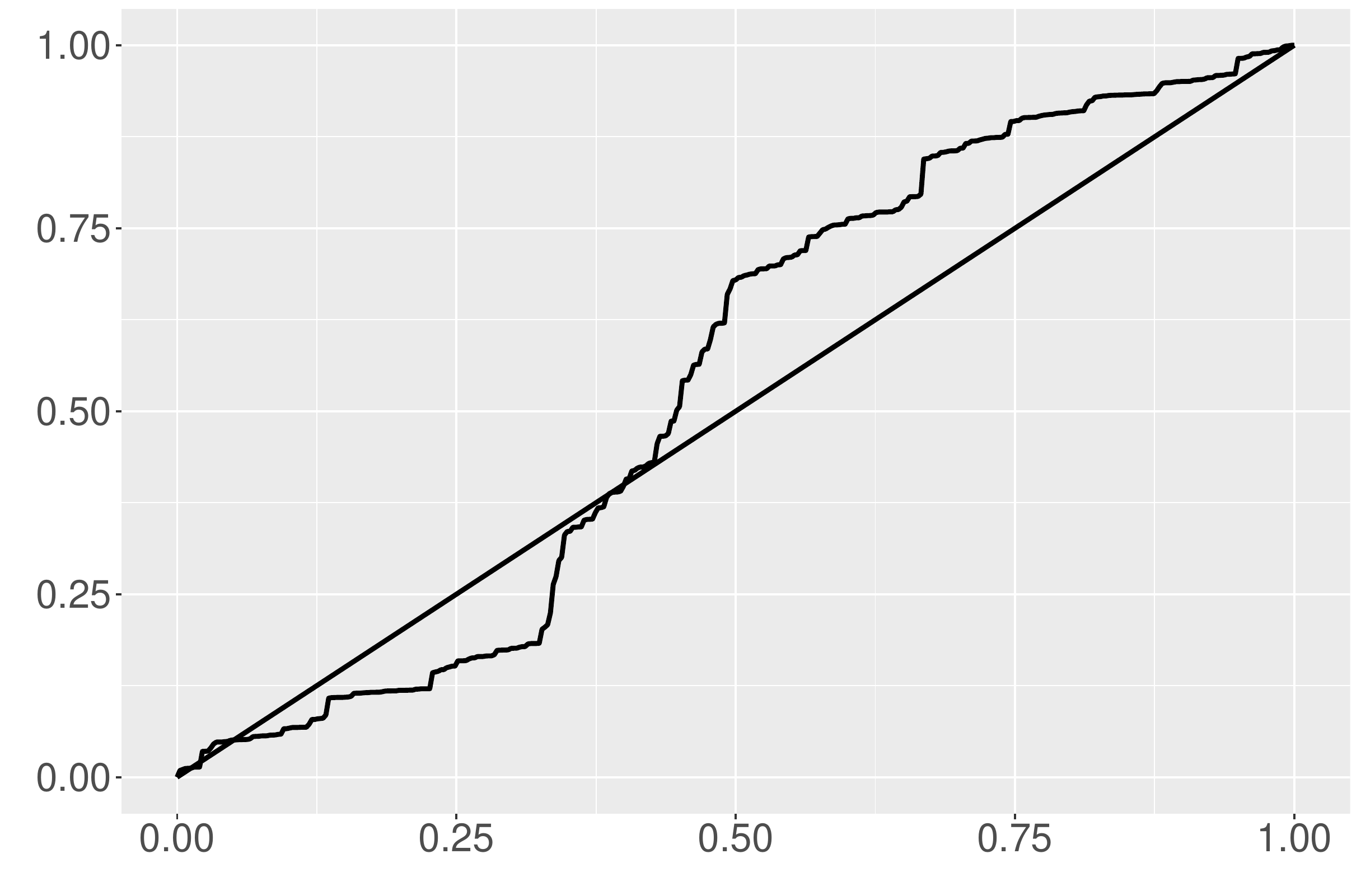}}
\subfigure[$\eta_{xrp}$]{\includegraphics[width=0.31\linewidth]{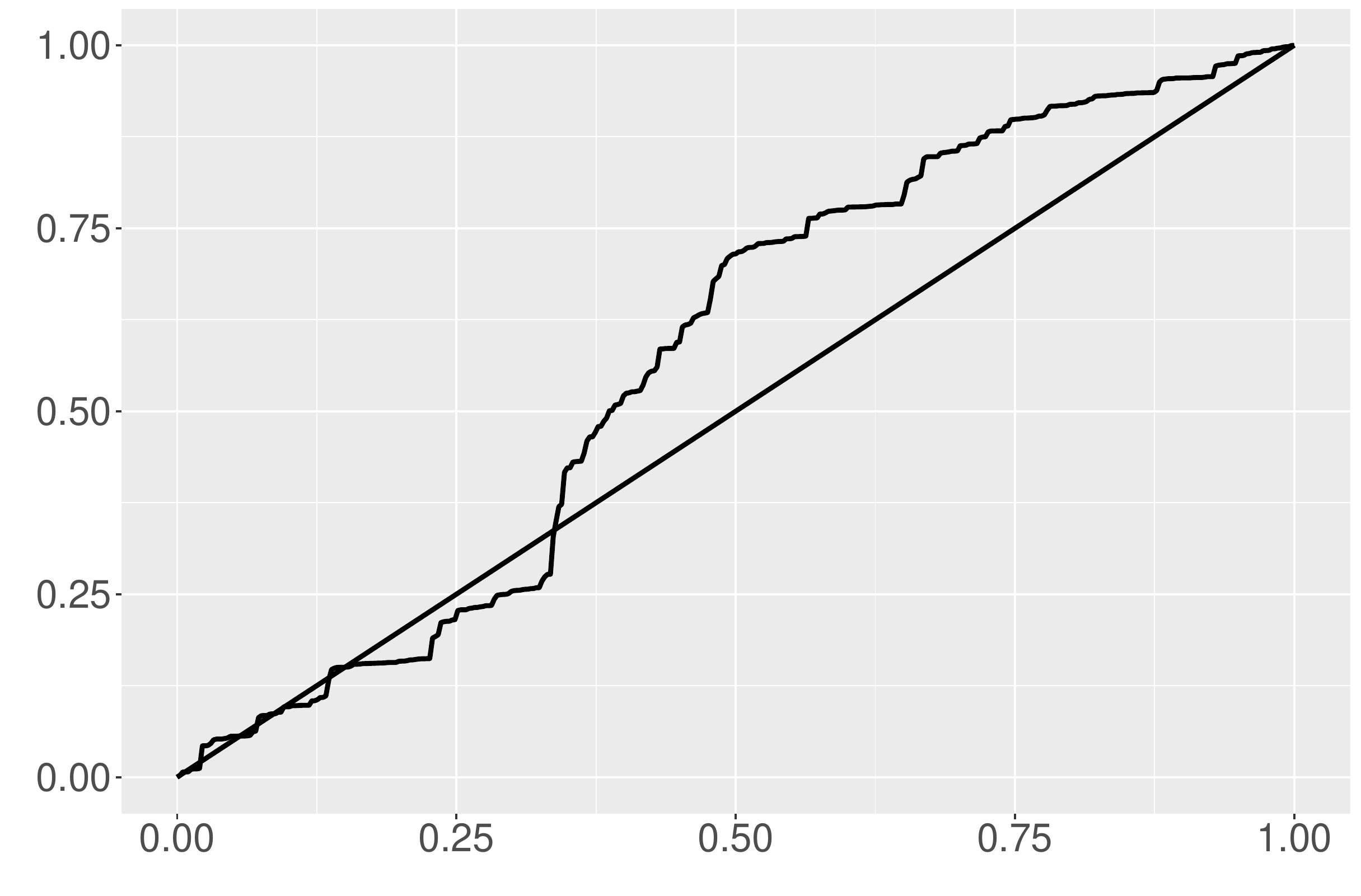}}\\
\subfigure[xlm]{\includegraphics[width=0.31\linewidth]{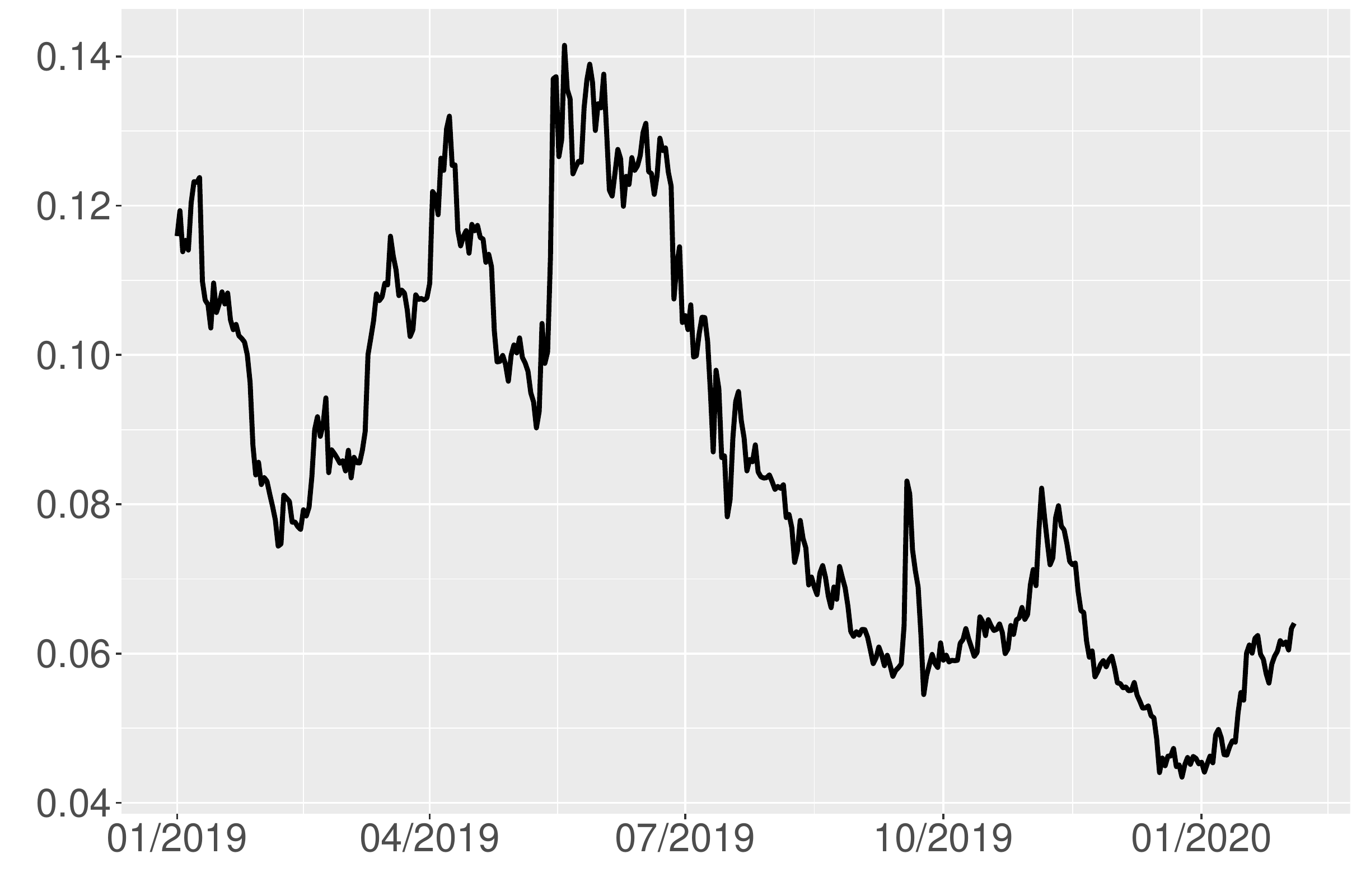}}
\subfigure[bch]{\includegraphics[width=0.31\linewidth]{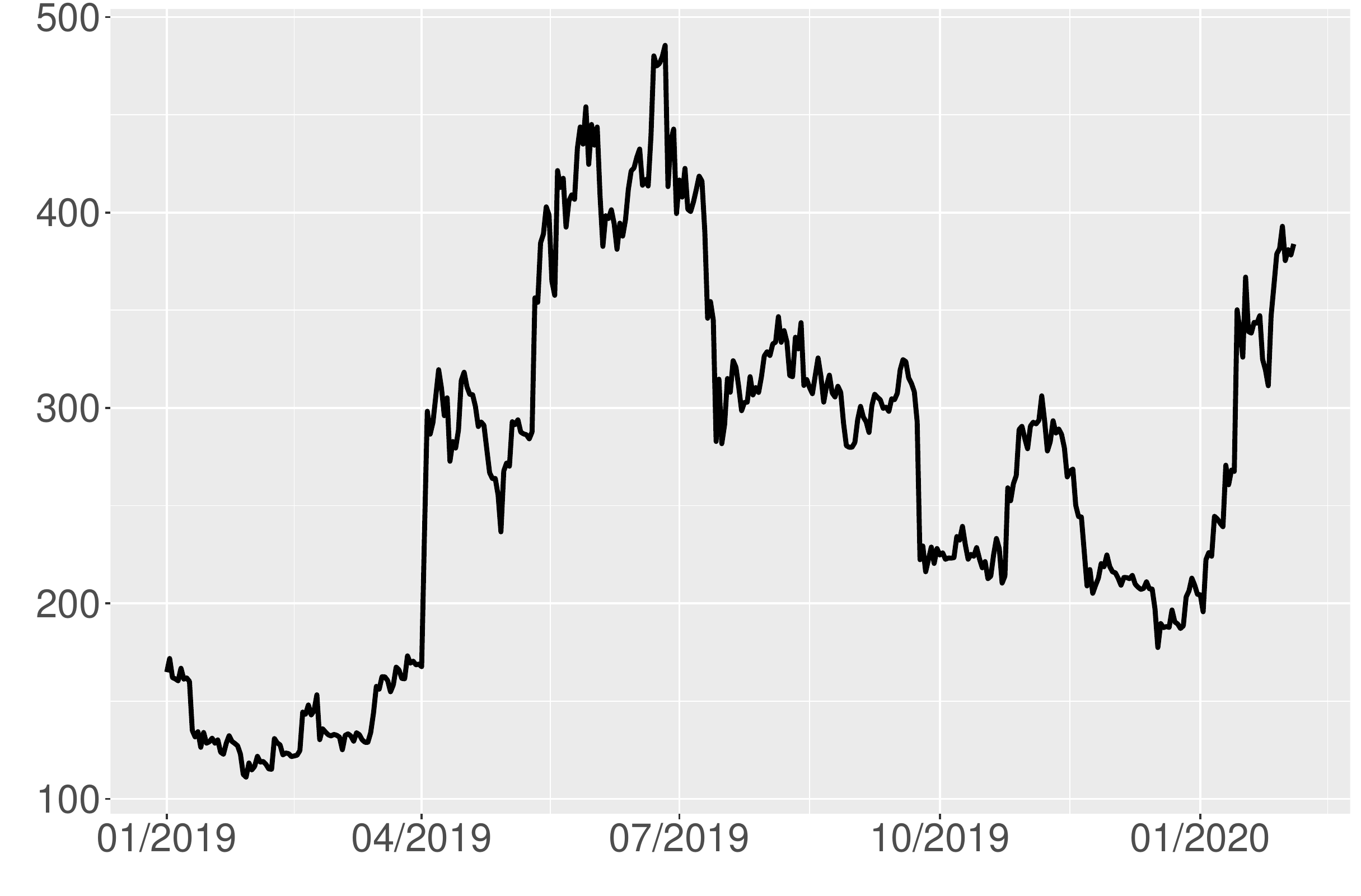}}
\subfigure[ltc]{\includegraphics[width=0.31\linewidth]{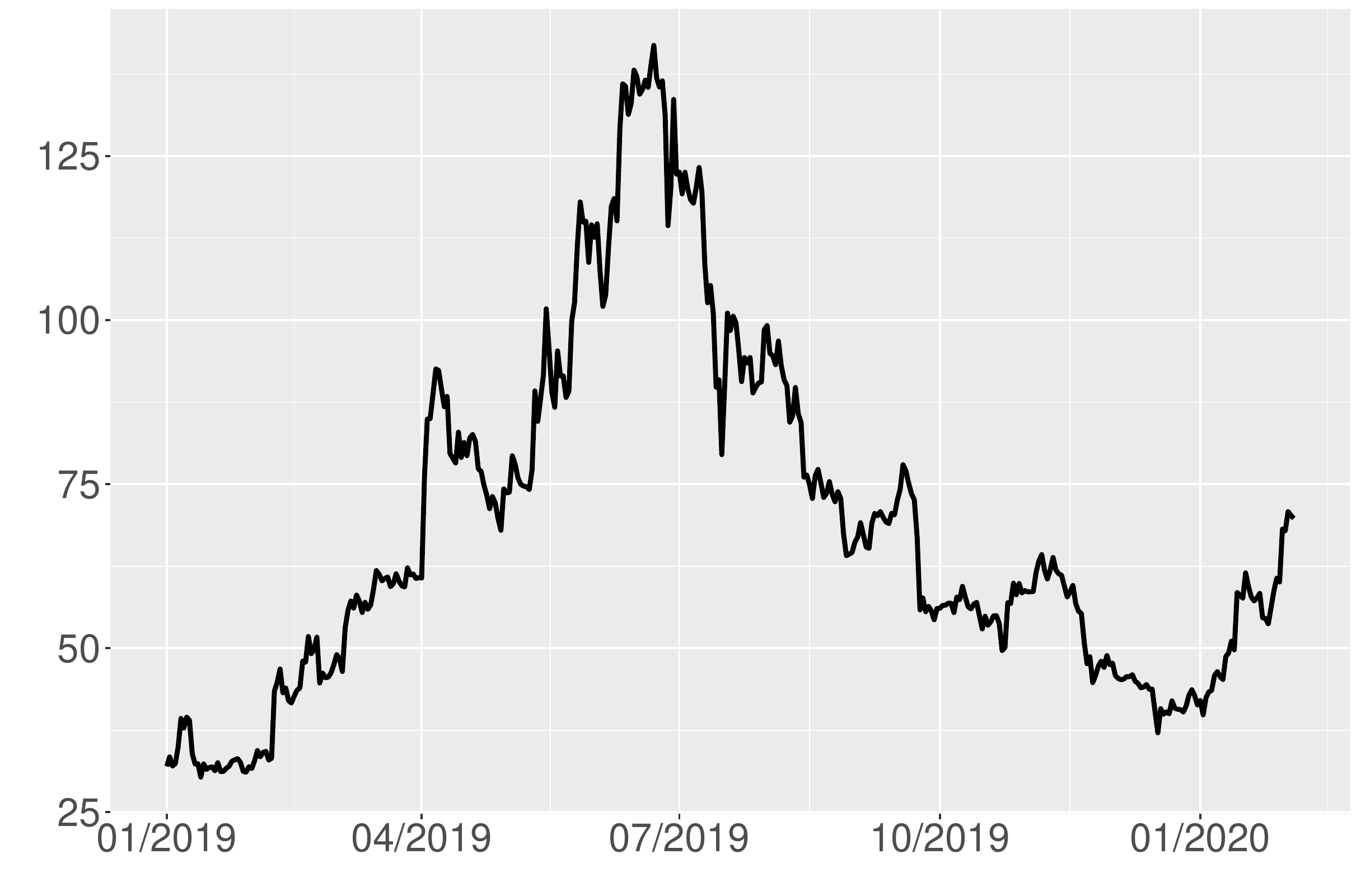}}
\subfigure[$\eta_{xlm}$]{\includegraphics[width=0.31\linewidth]{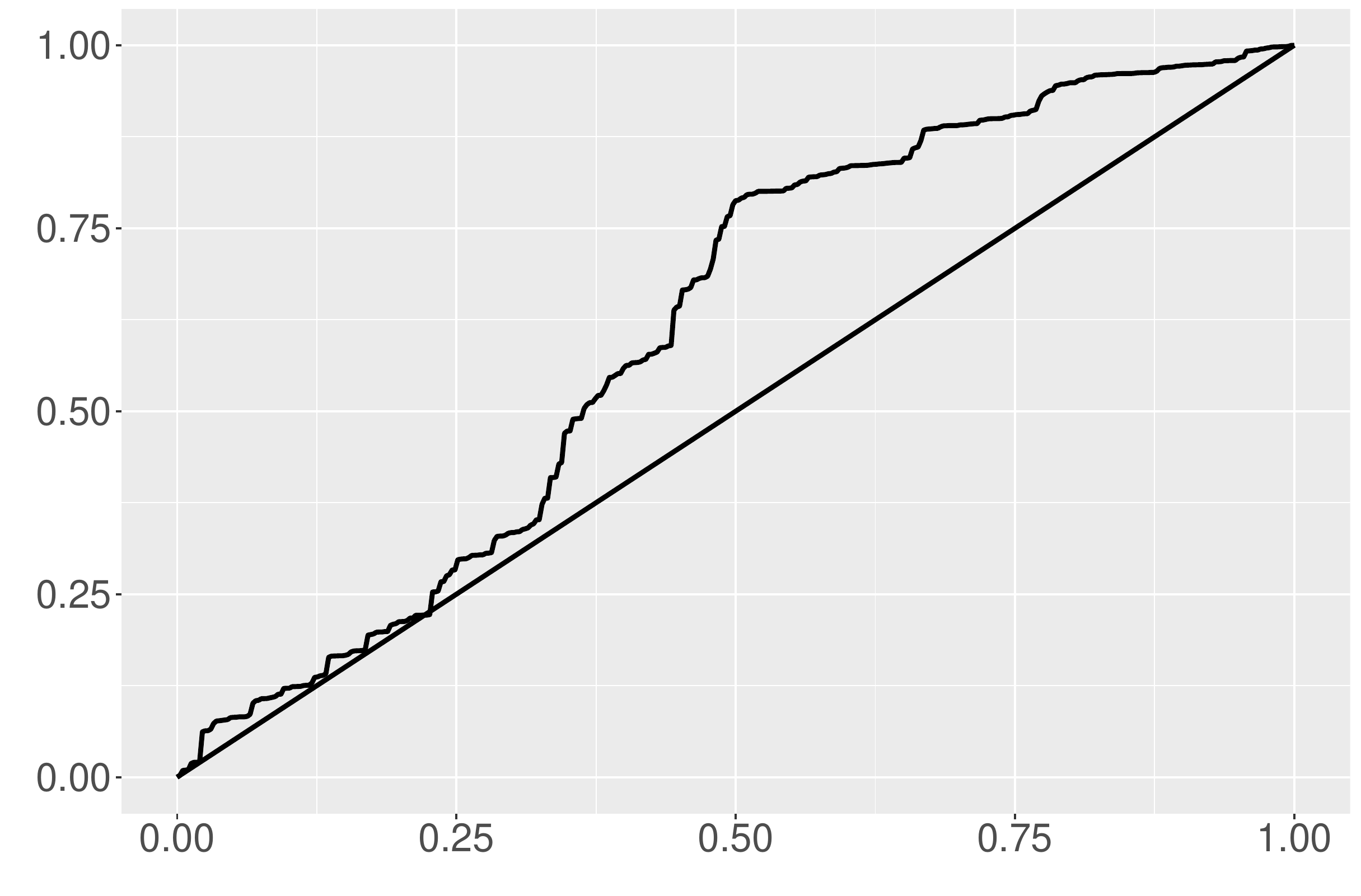}}
\subfigure[$\eta_{bch}$]{\includegraphics[width=0.31\linewidth]{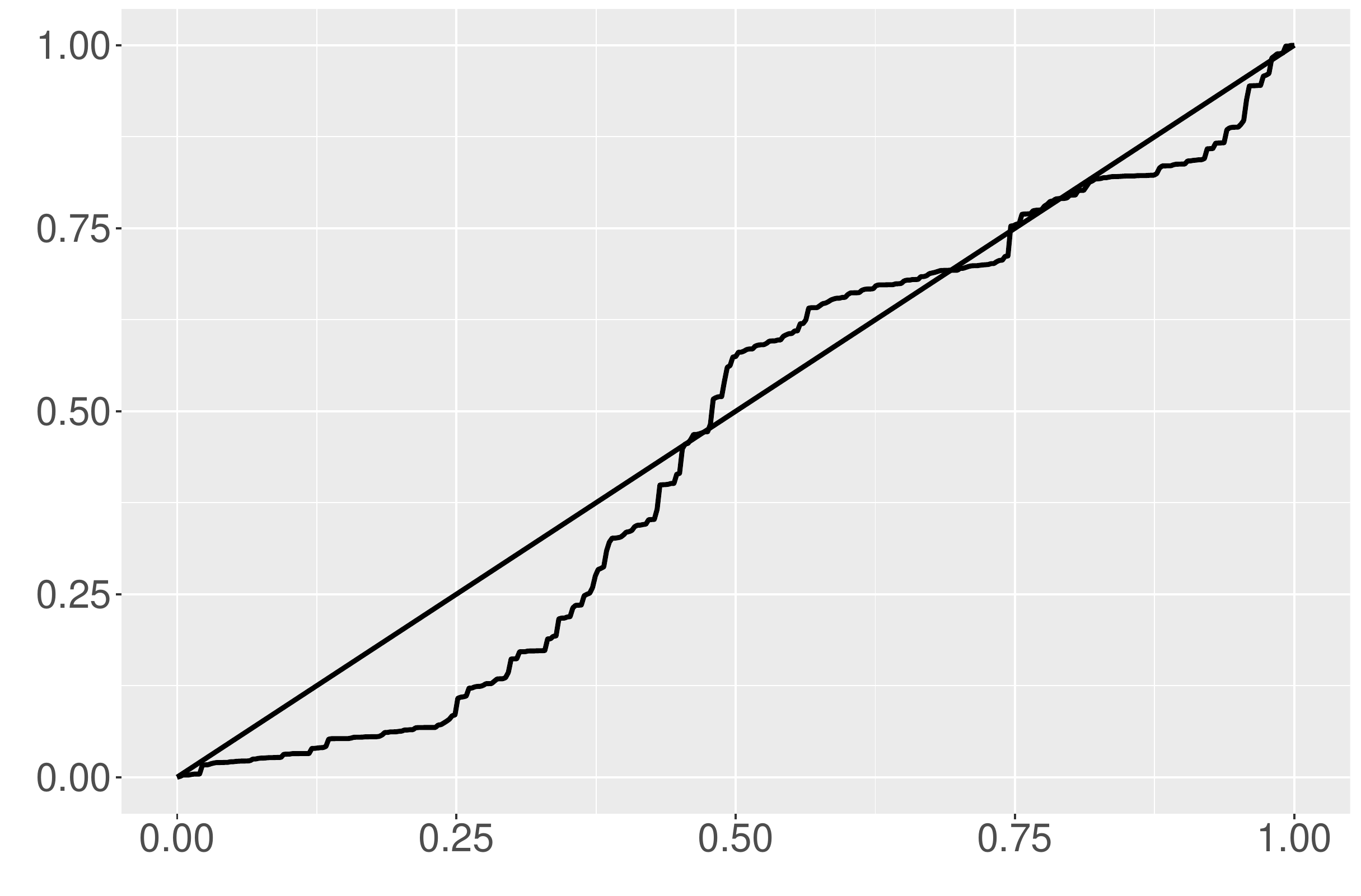}}
\subfigure[$\eta_{ltc}$]{\includegraphics[width=0.31\linewidth]{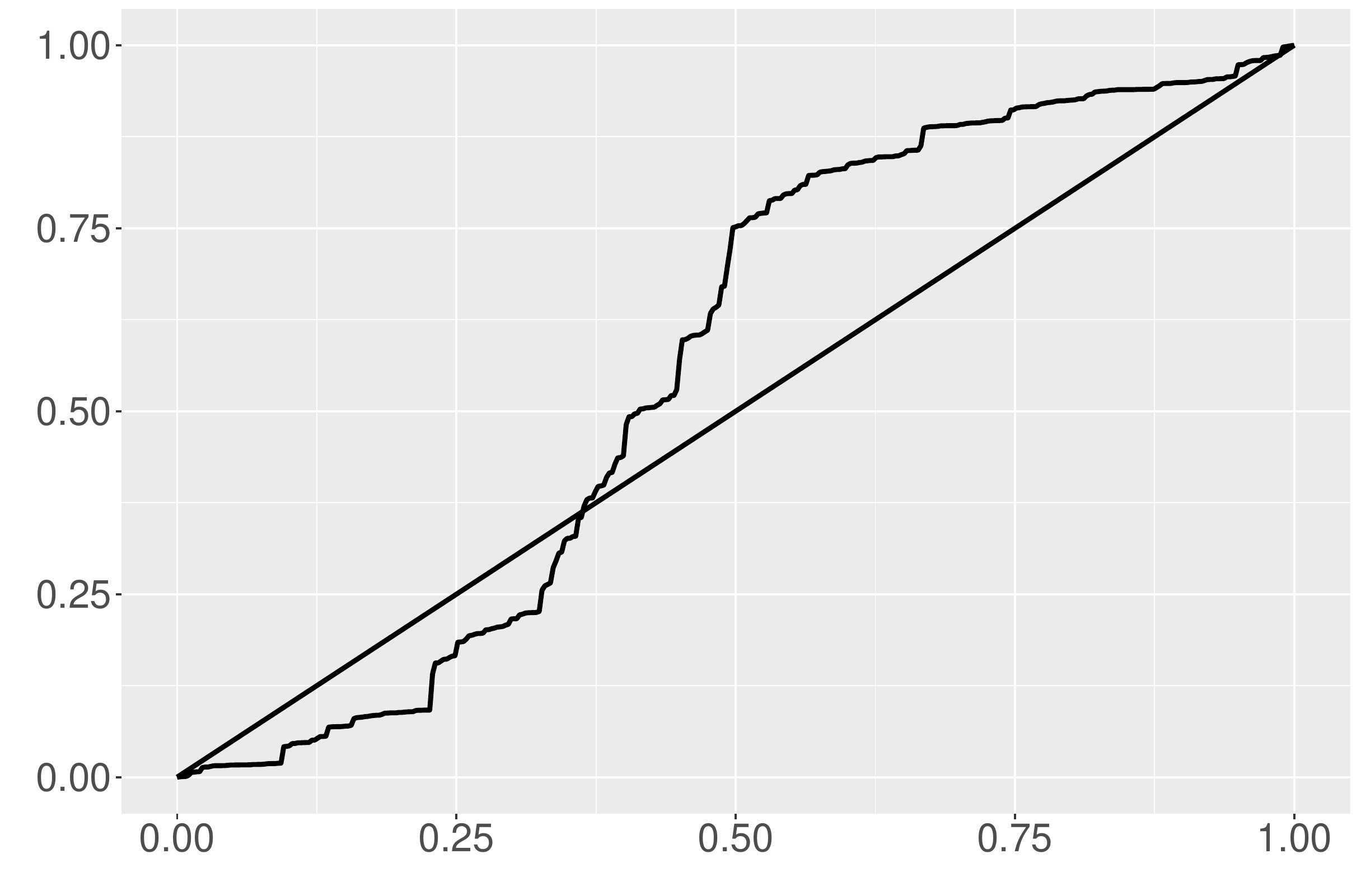}}\\
\end{center}%
\caption{Time series plot of btc, eth, xrp, xlm, bch, ltc and corresponding variance profiles}
\label{fig1}
\end{figure}

\begin{figure}[h]%
\begin{center}%
\subfigure[eos]{\includegraphics[width=0.31\linewidth]{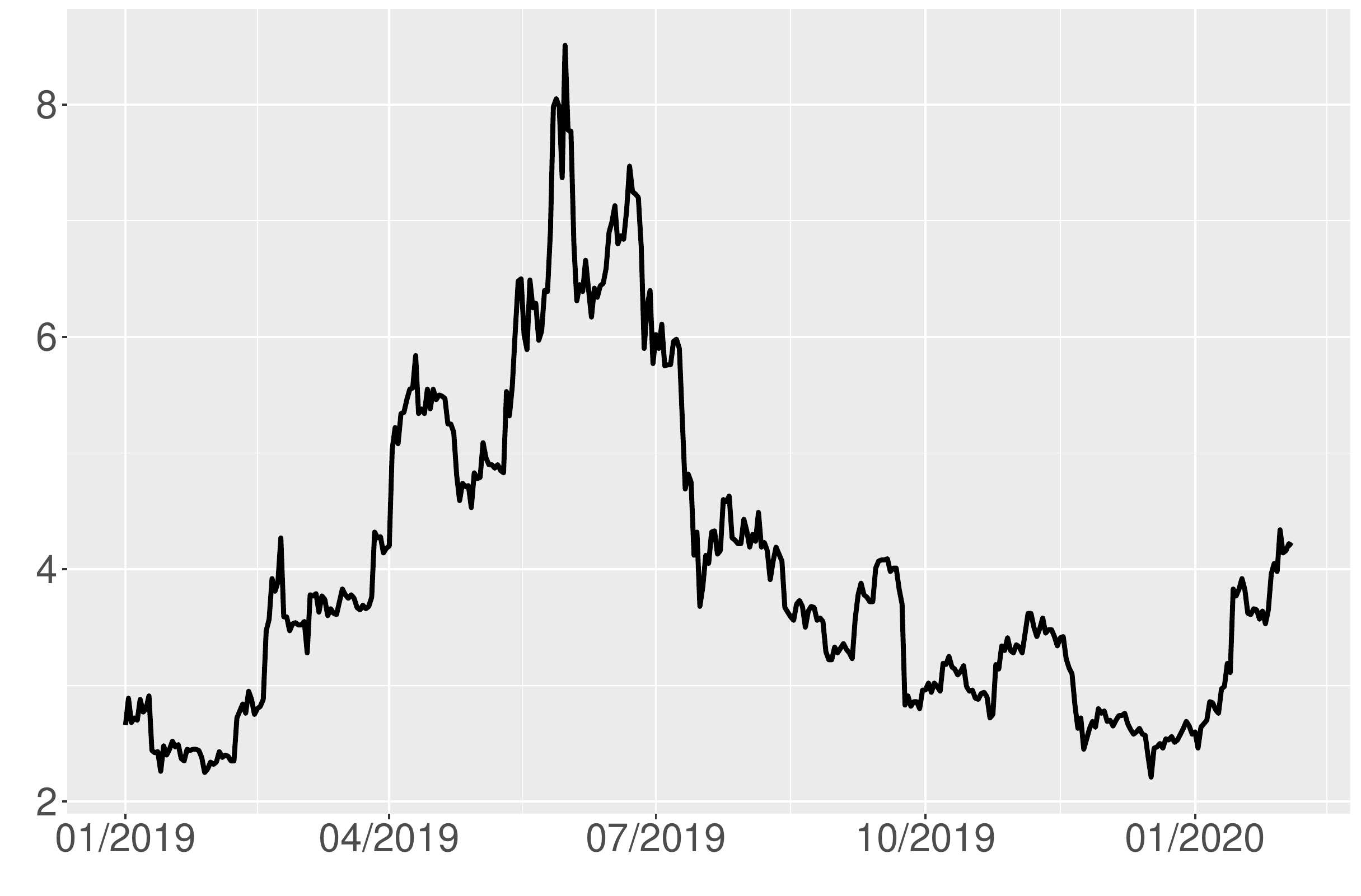}}
\subfigure[bnb]{\includegraphics[width=0.31\linewidth]{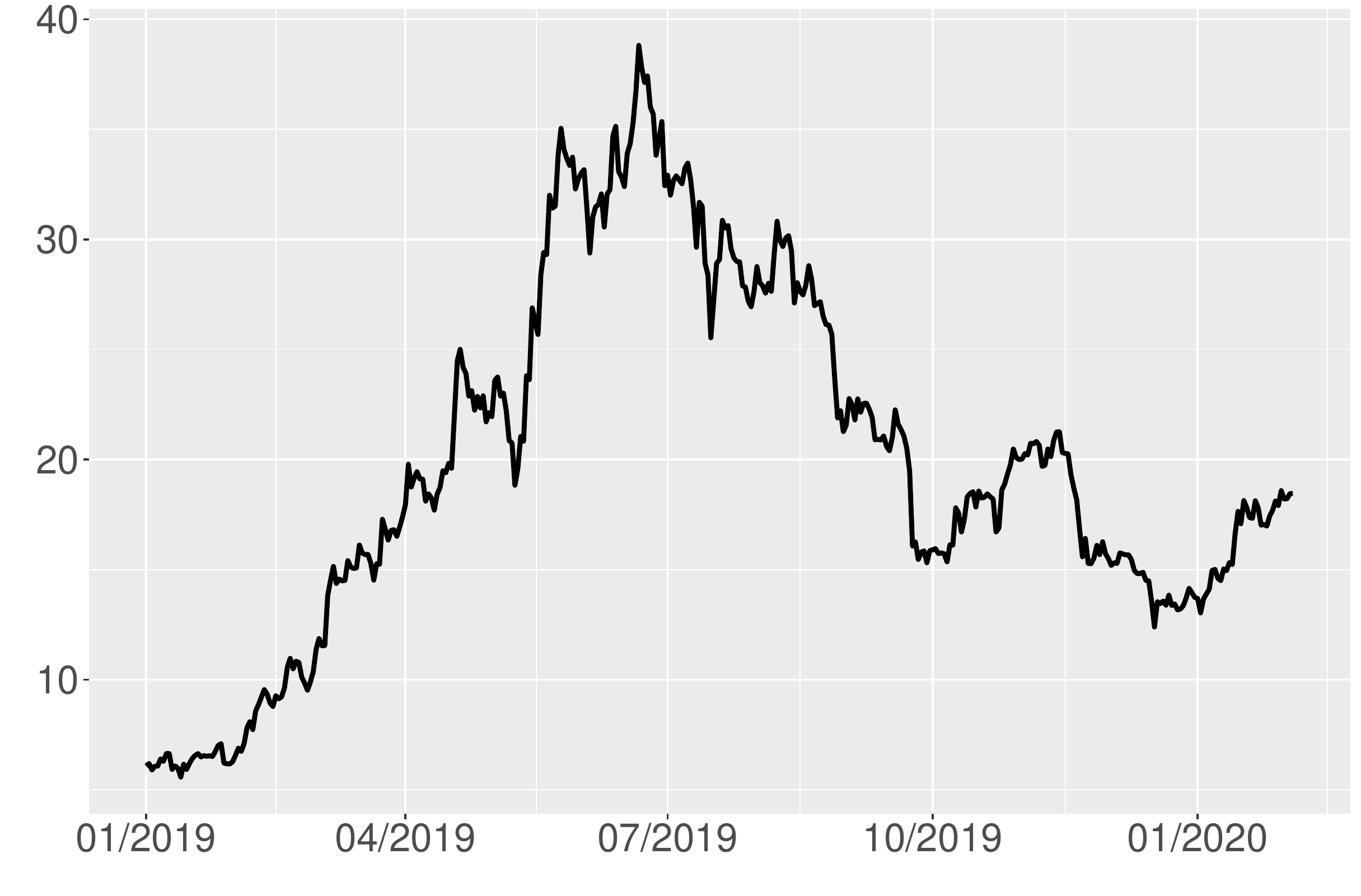}} 
\subfigure[ada]{\includegraphics[width=0.31\linewidth]{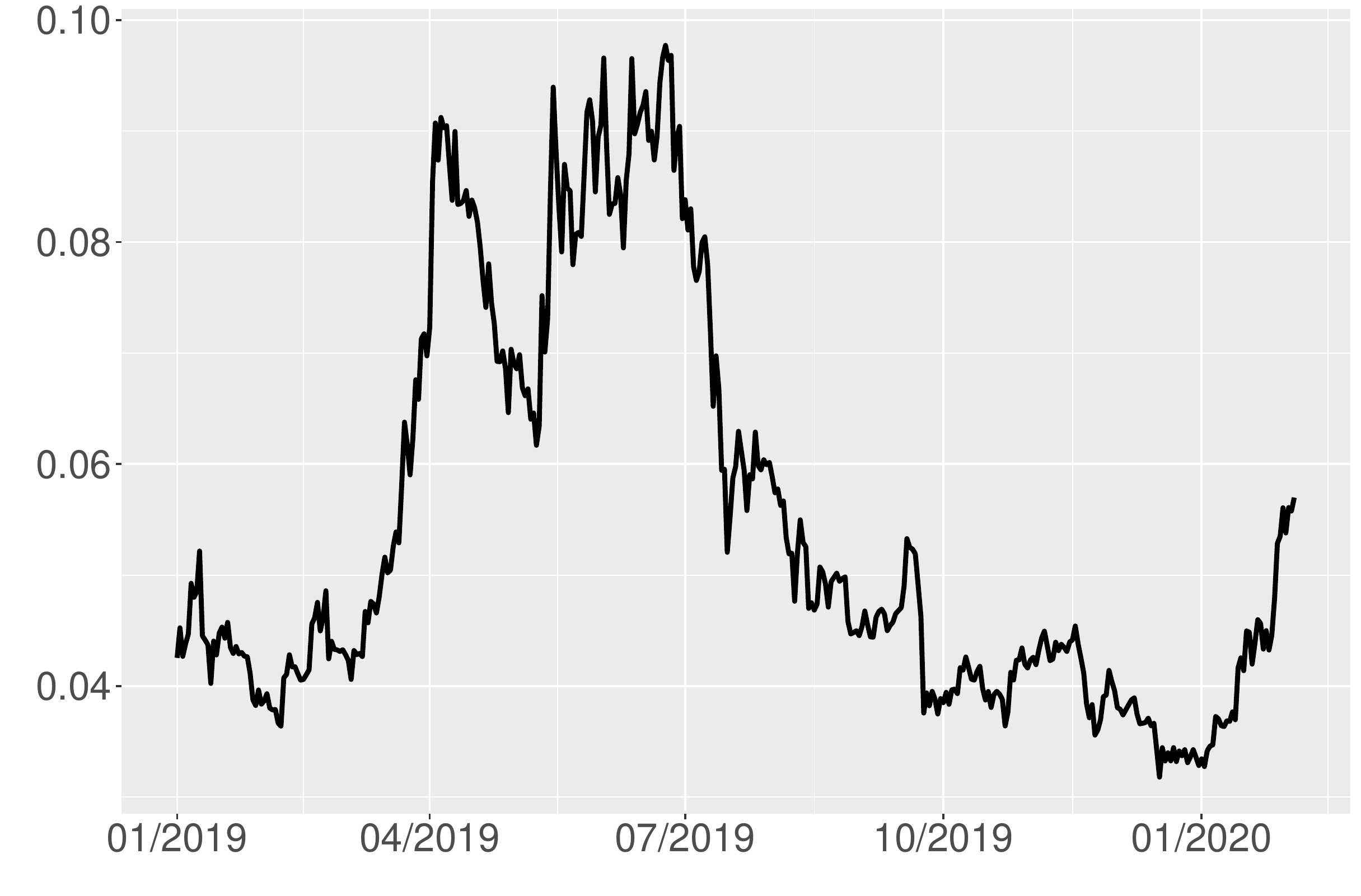}}
\subfigure[$\eta_{eos}$]{\includegraphics[width=0.31\linewidth]{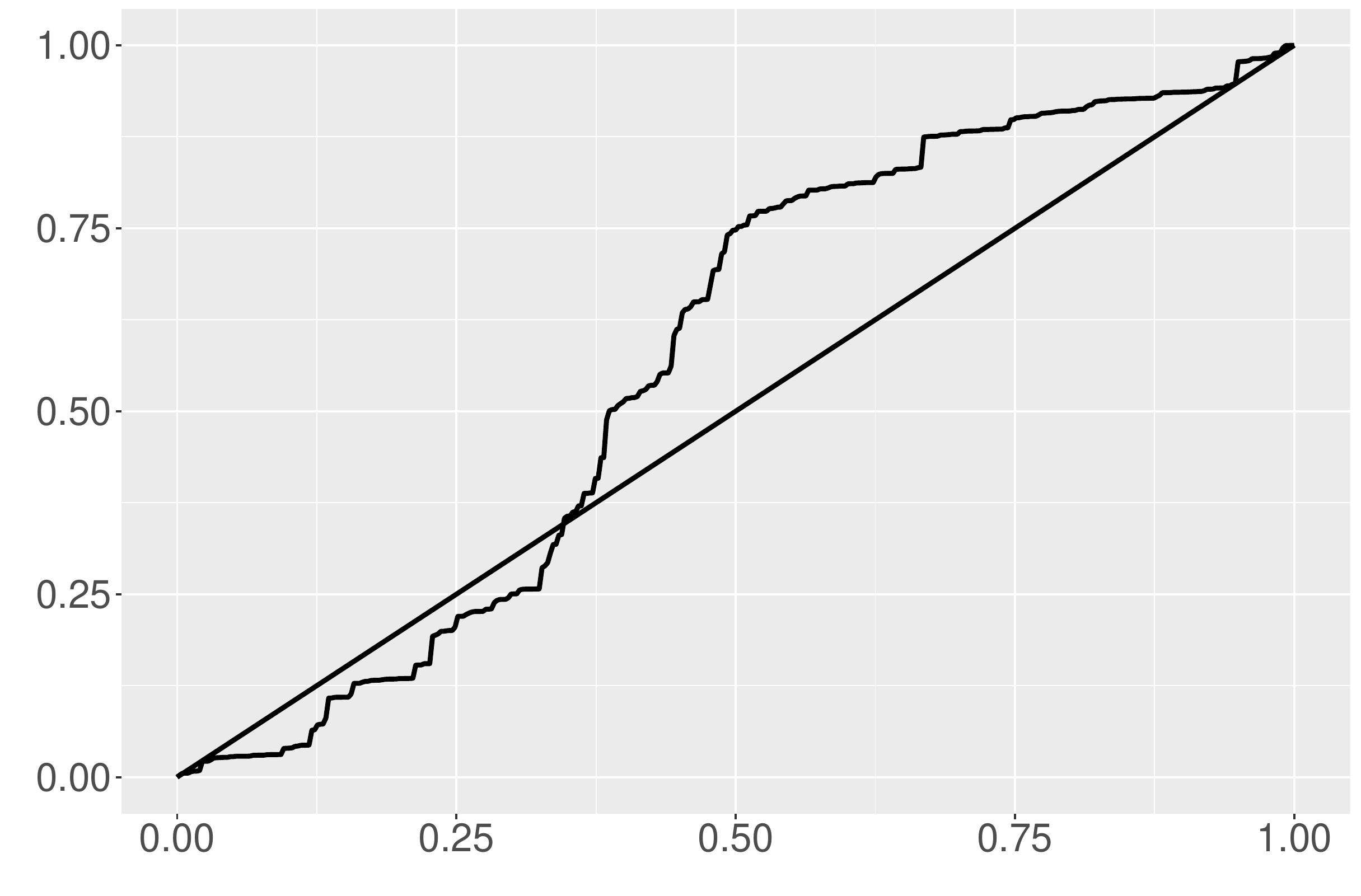}}
\subfigure[$\eta_{bnb}$]{\includegraphics[width=0.31\linewidth]{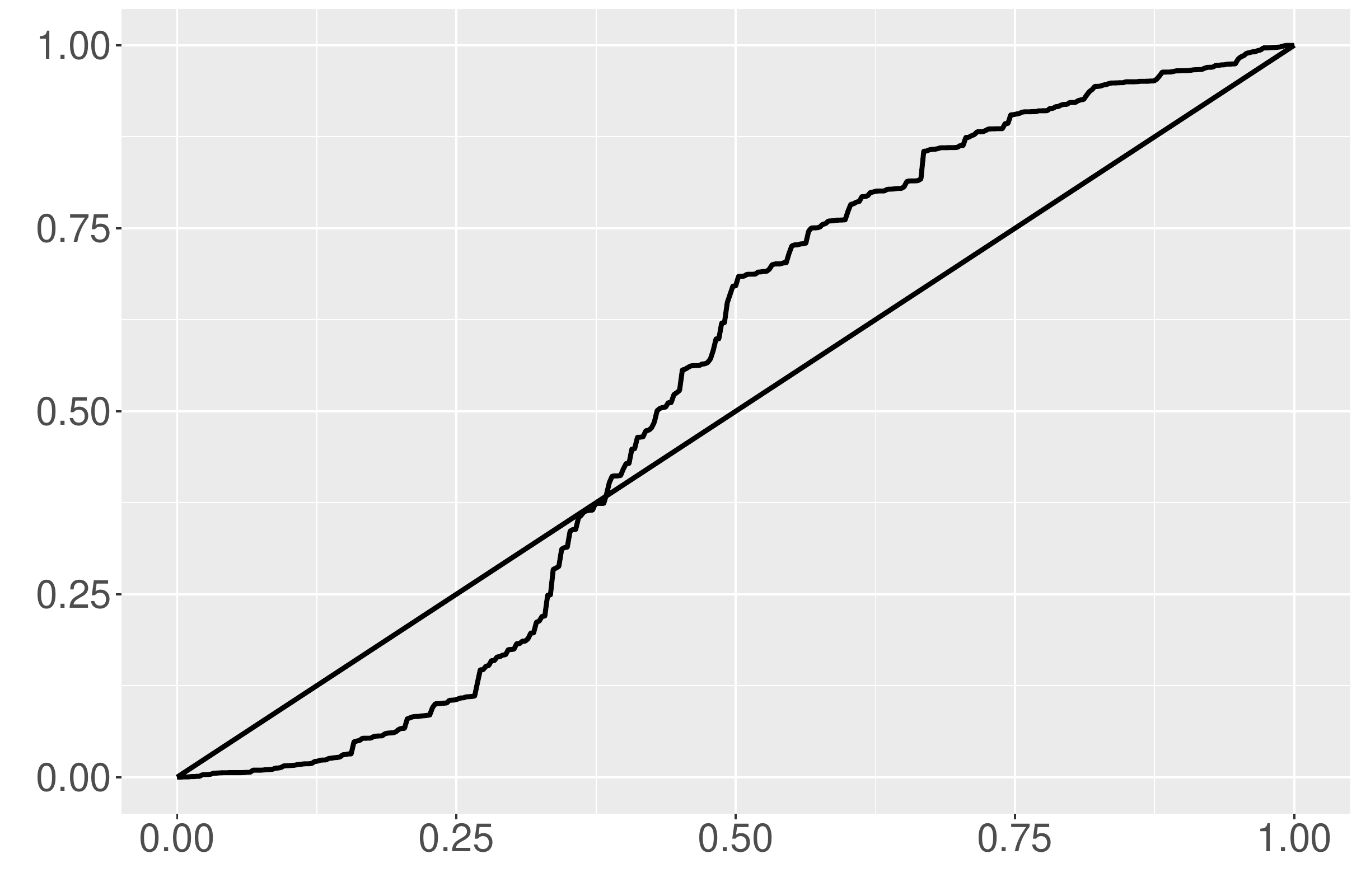}}
\subfigure[$\eta_{ada}$]{\includegraphics[width=0.31\linewidth]{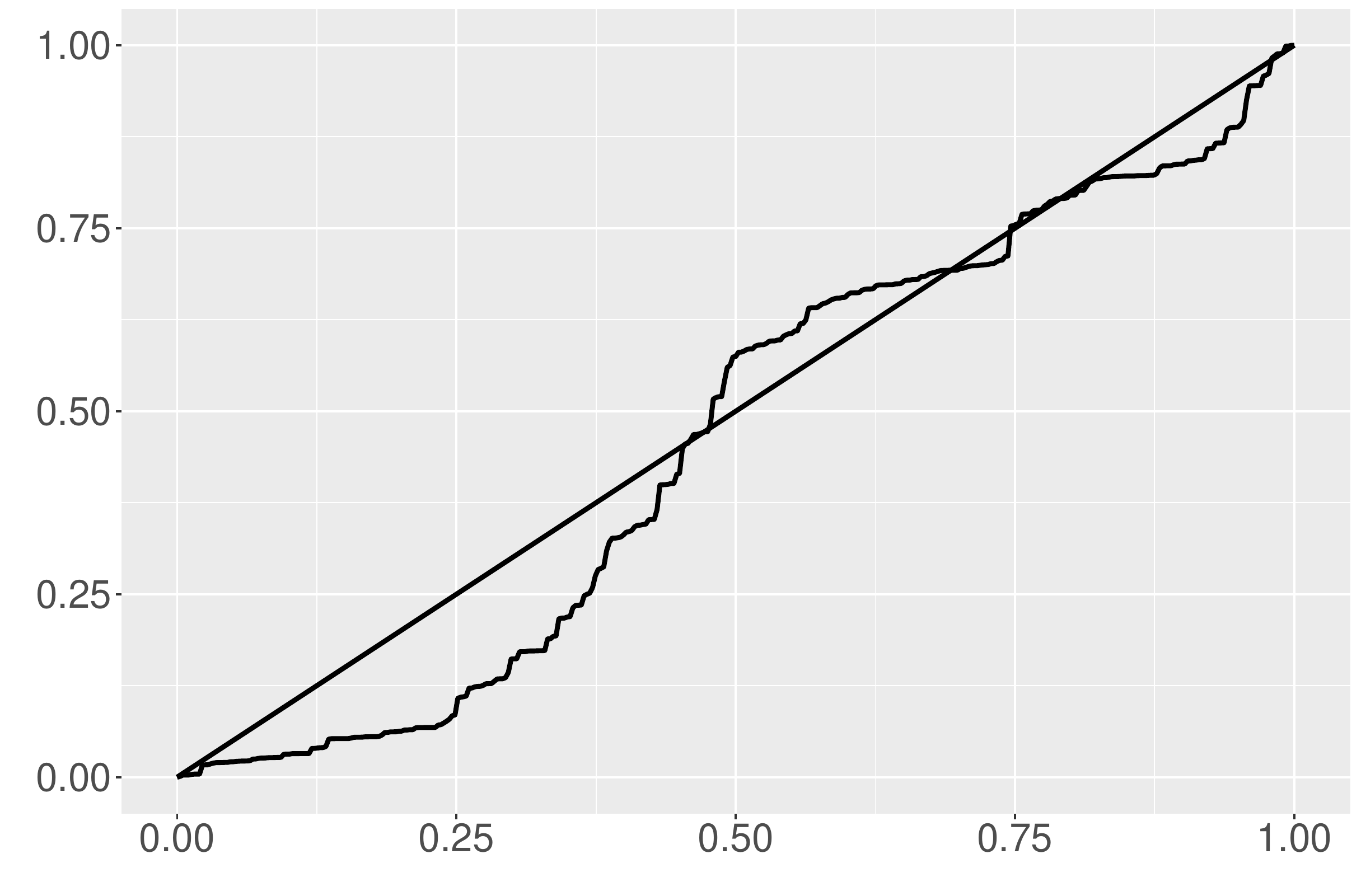}}\\
\subfigure[xtz]{\includegraphics[width=0.31\linewidth]{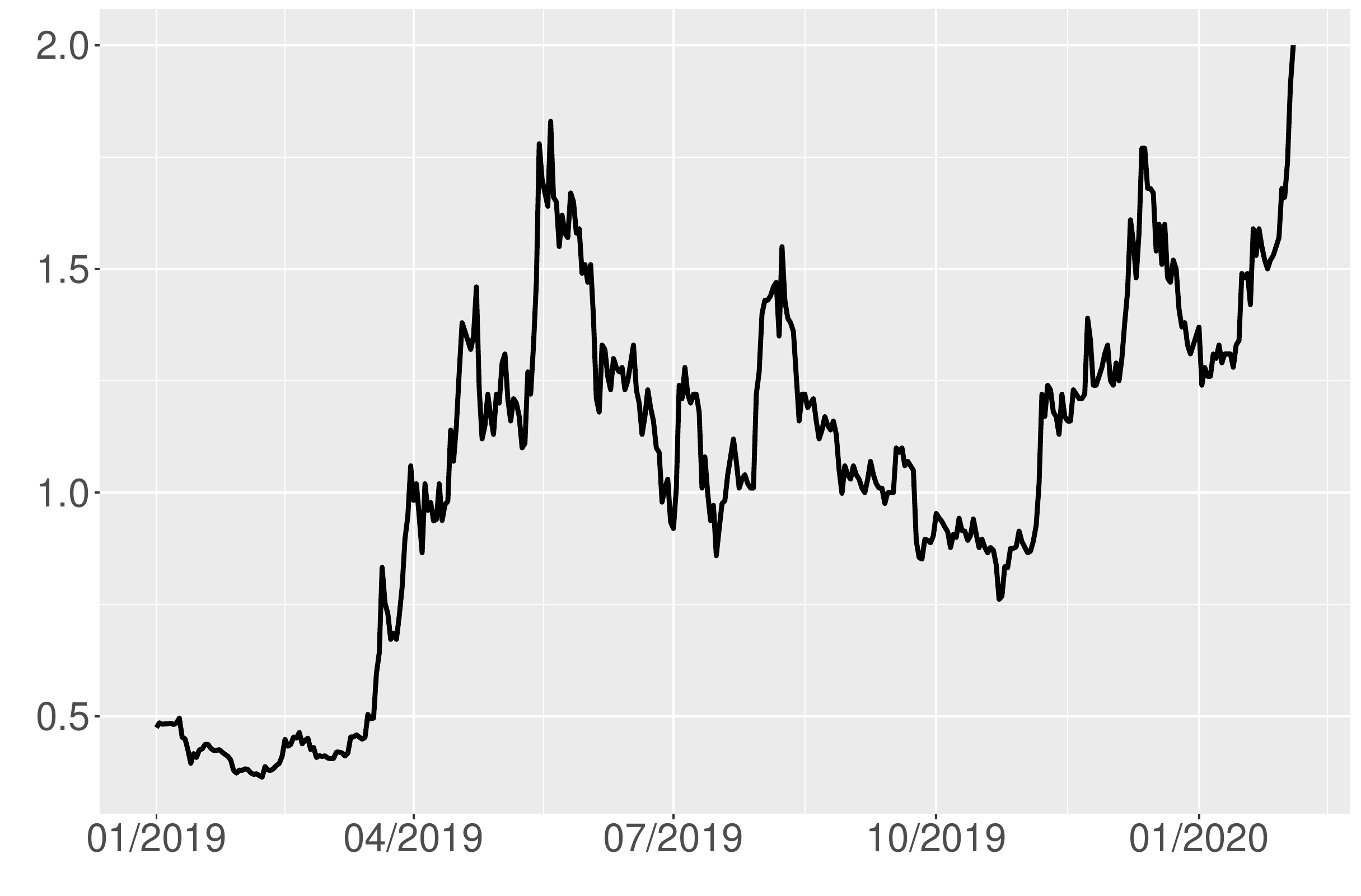}}
\subfigure[etc]{\includegraphics[width=0.31\linewidth]{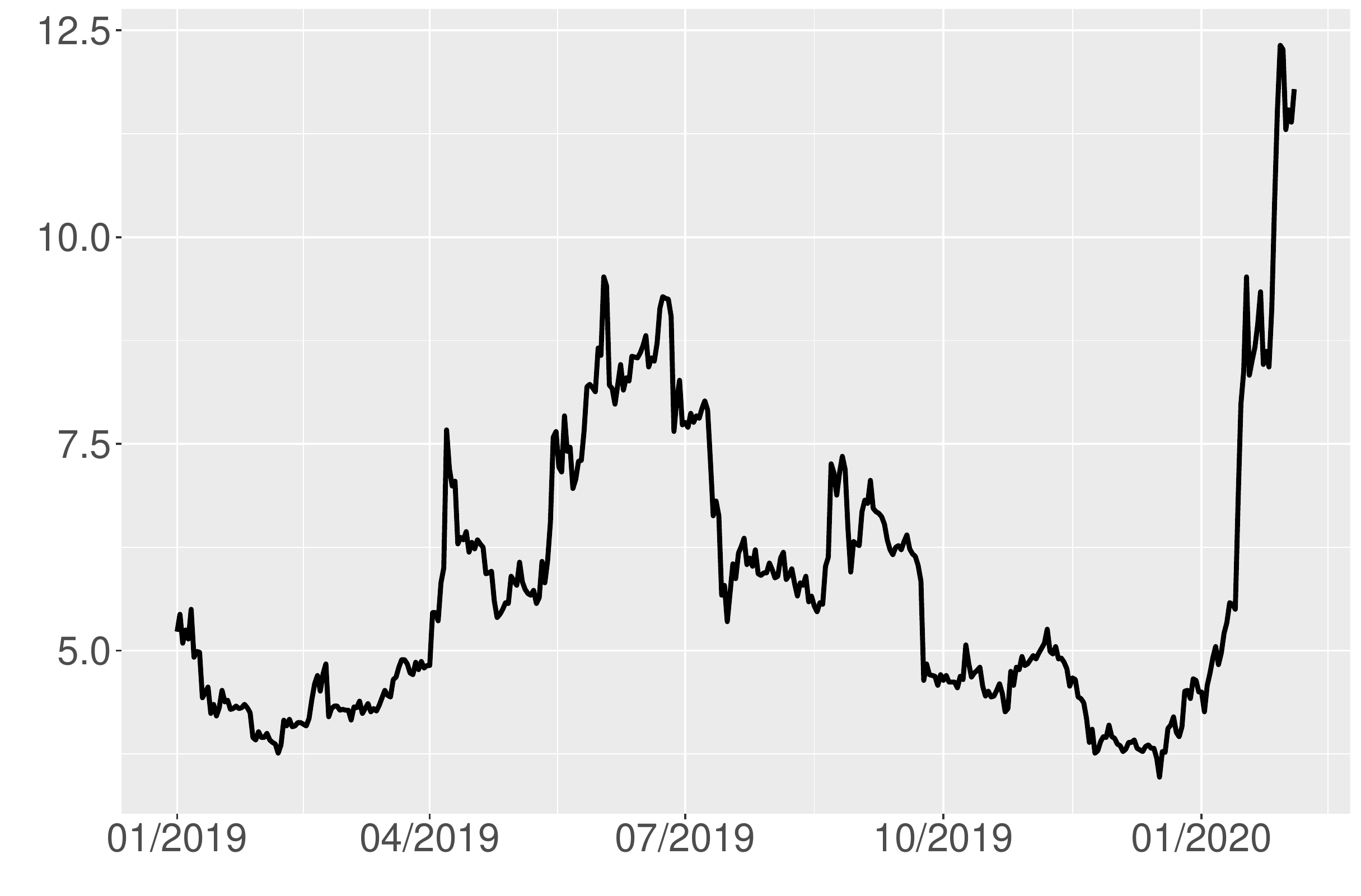}}
\subfigure[xmr]{\includegraphics[width=0.31\linewidth]{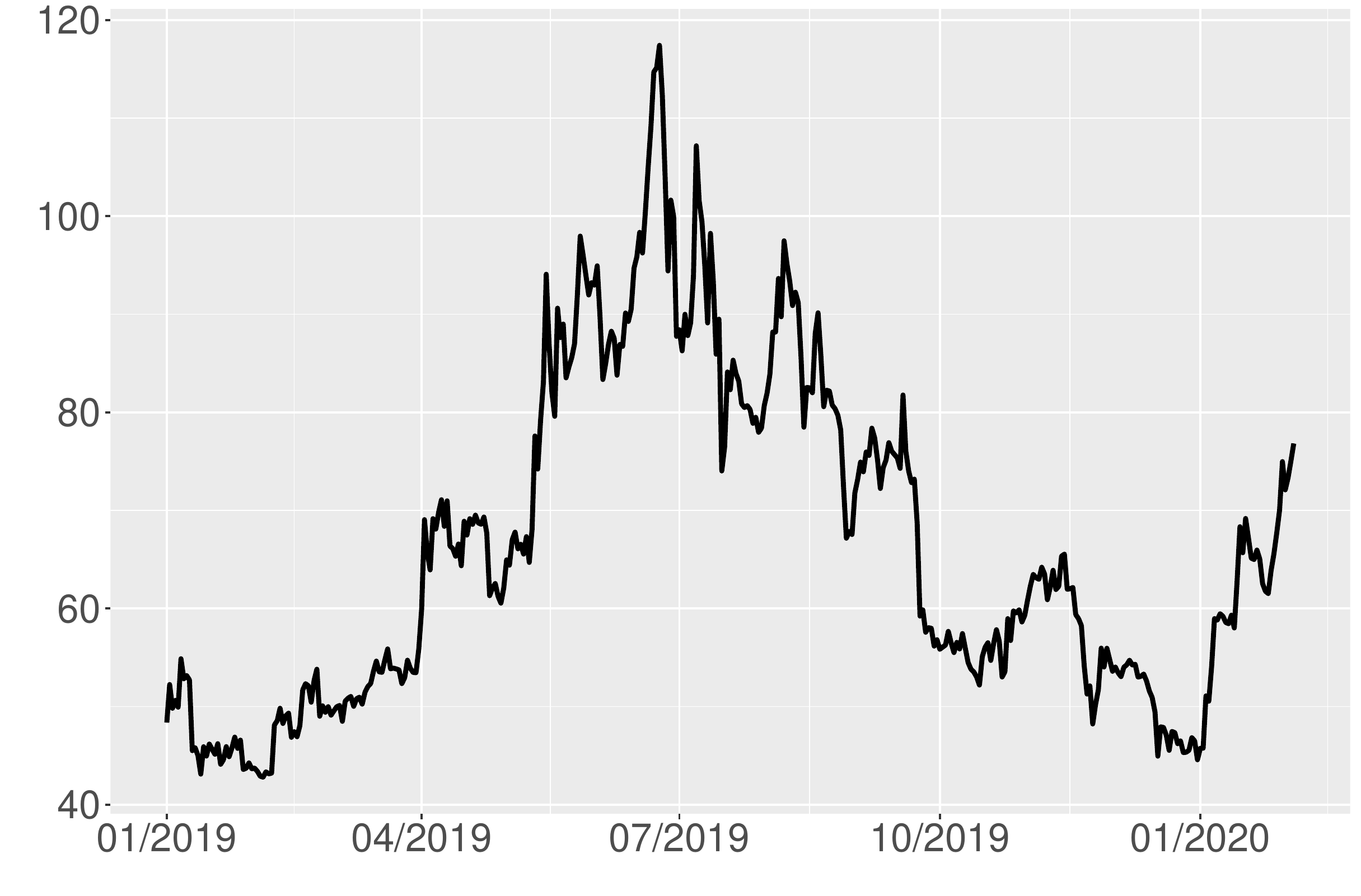}}
\subfigure[$\eta_{xtz}$]{\includegraphics[width=0.31\linewidth]{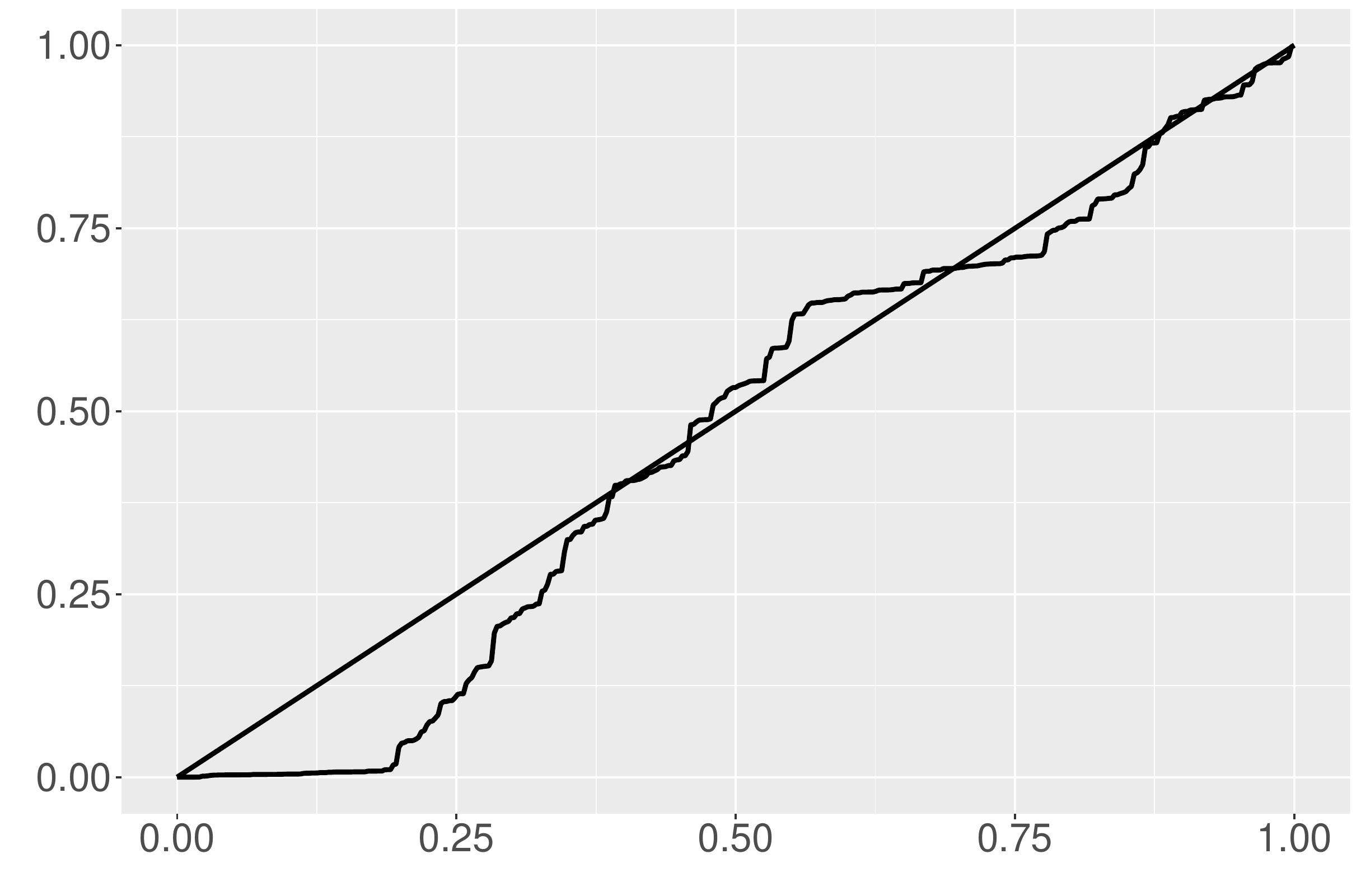}}
\subfigure[$\eta_{etc}$]{\includegraphics[width=0.31\linewidth]{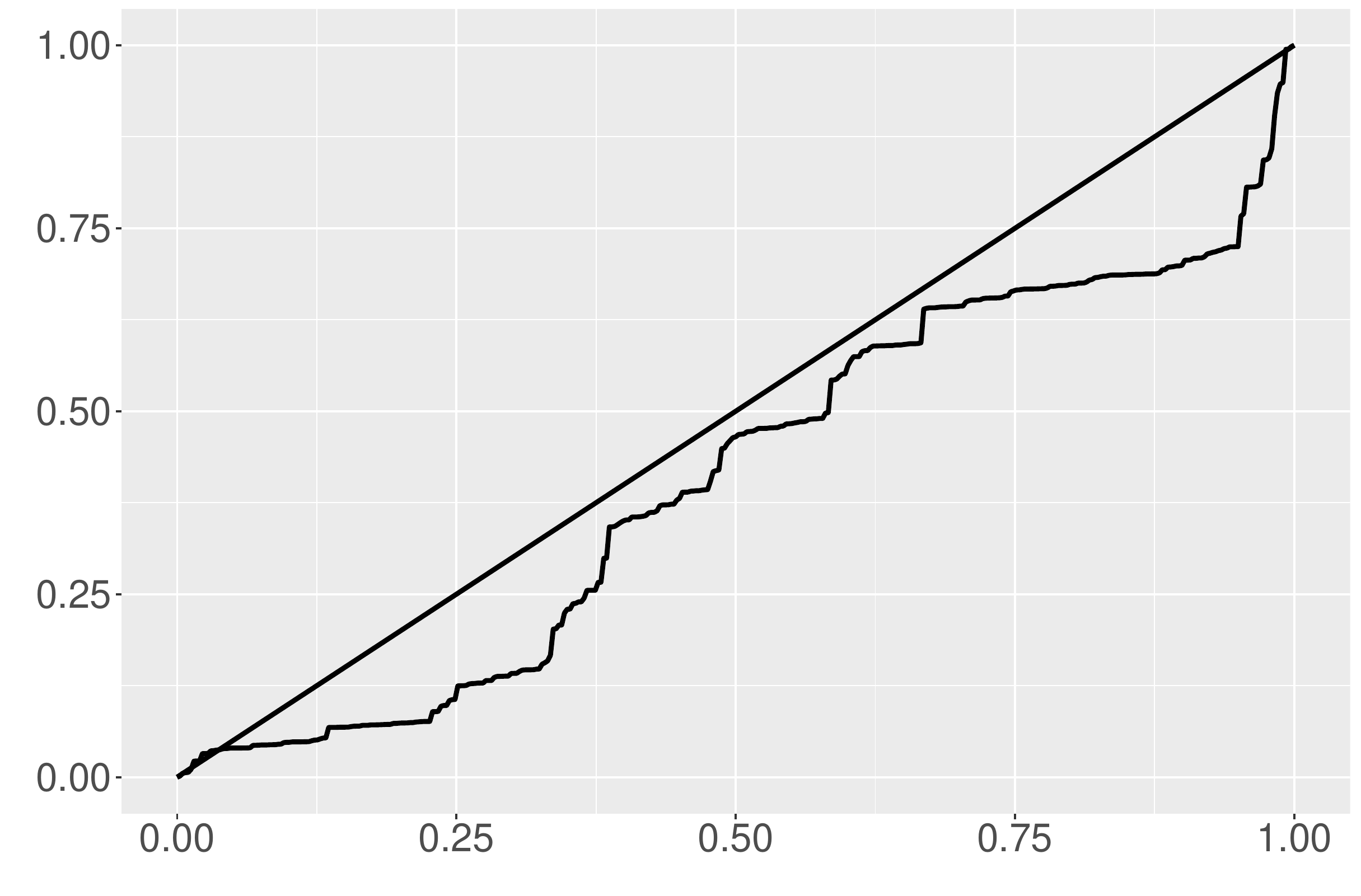}}
\subfigure[$\eta_{xmr}$]{\includegraphics[width=0.31\linewidth]{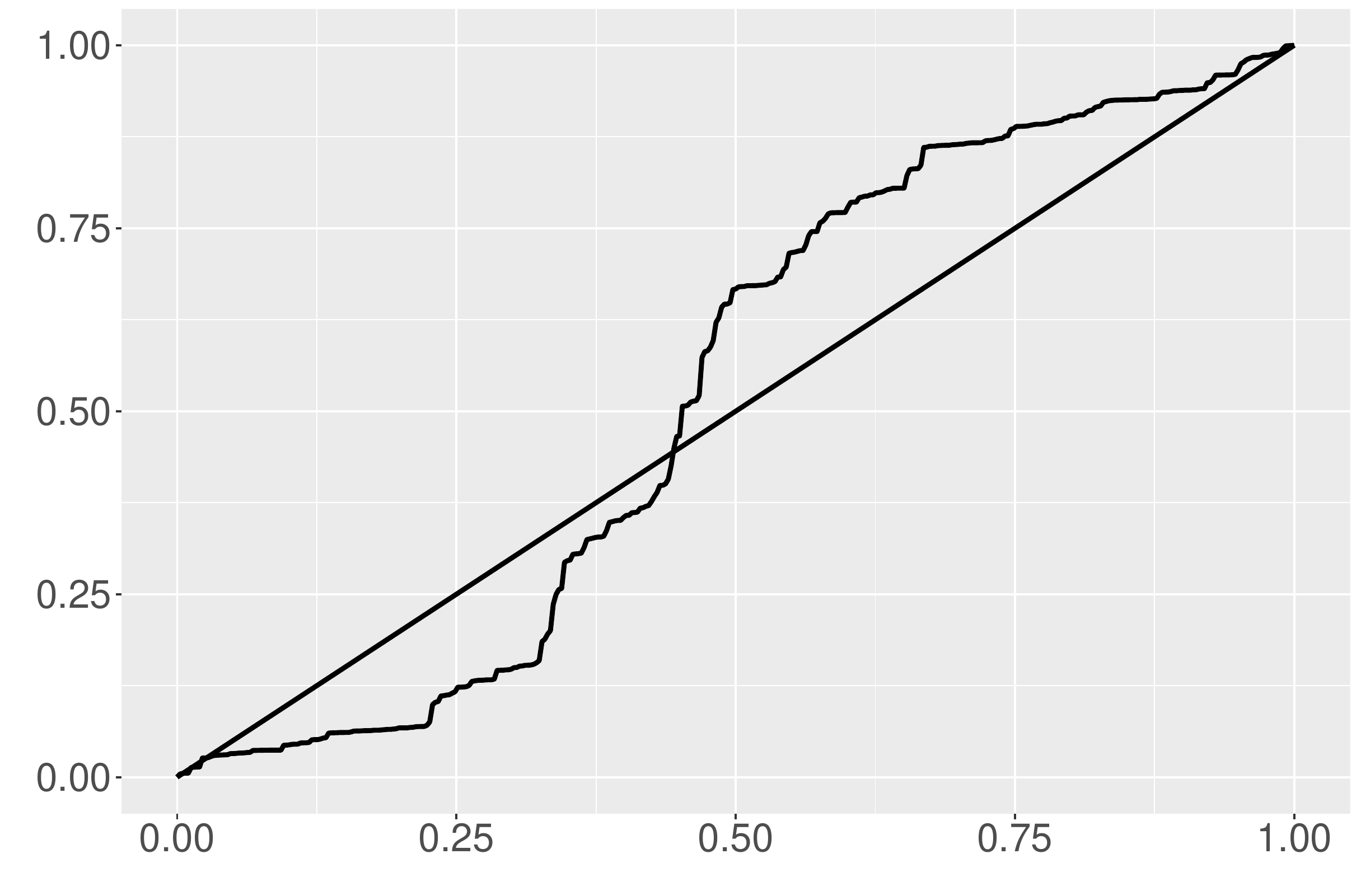}}\\
\end{center}%
\caption{Time series plot of eos, bnb, ada, xtz, etc, xmr and corresponding variance profiles}
\label{fig2}
\end{figure}

In this section, we investigate whether or not we observe the emergence of a bubble in the top 12 largest cryptocurrencies by capitalization as of February 3, 2020 (btc, eth, xrp, xlm, bch, ltc, eos, bnb, ada, xtz, etc, xmr), from January 1, 2019 to February 3, 2020, which were 399 observations. In all cases, the closing price in US dollars at 00:00 GMT on the corresponding day is used. The time series with corresponding estimated variance profiles are plotted in Figures \ref{fig1} and \ref{fig2}. For comparative analysis, we use the same tests as in the previous section: the standard $SADF$ test, the wild bootstrap $SADF$ test ($SADF_b$), a union of rejections of the $SADF_b$ and SBZ tests ($SBZ$), a union of rejections of the $SADF_b$ and sign-based tests ($sSADF$) and the $STADF$ test. For the wild bootstrap $p$-values, $B = 999$ bootstrap replications were used. For the standard $SADF$ test and the time-transformed test $STADF$, the $p$-values are obtained by simulations of the asymptotic distributions of the test statistics under homoskedasticity. Table \ref{tab_emp} presents the results of the tests (p-values) for the time series considered. We use $r_0=0.1$ for ease of calculations of the p-values; the overall tendency will be the same even if we use $r_0= \lfloor 0.01+1.8/\sqrt{T} \rfloor$. 

From Table \ref{tab_emp}, we observe several patterns of the $p$-values. For btc and ada series, the $SADF$ strongly rejects the null hypothesis and the other robust tests also reject the null but with larger $p$-values. This implies that we observe a bubble in this period for btc and ada series but the upsurge behaviour in the sample is partially explained by the volatility change. For eth, bch, ltc, eos, bnb, xtz and etc series, the p-values of the SADF test is below 0.05 or 0.1 but the other robust tests, except for $SADF_b$ for eth, bch and xtz and $SBZ$ for bnb and xtz, do not reject the null hypothesis at the 10\% level. Note that the variance profiles of these series show that there exist some regimes in which volatility moves from low to high. In this case, the SADF test suffers from strong over-rejection and also the $SADF_b$ and $SBZ$ tests tend to be mildly over-sized as is observed in the simulation section. By taking this into account, we should not reject the null hypothesis and there is no strong evidence of the existence of a bubble for these series. Finally, for xrp, xlm and xmr series, we do not observe the existence of a bubble by all the tests.

As demonstrated above, we should carefully interpret the empirical results of the tests for a bubble using the estimated variance profile. If we observe a regime with the upward volatility shift, the standard test as well as some of the robust tests tend to over-reject the null hypothesis. The inspection of the combination of the bubble tests and the variance profile will lead us to a more reliable result.

Finally, Table \ref{tab_emp2} demonstrates the  computation time of $SADF$, $SADF_b$ and $STADF$ for long time series with thousands observations. It can be seen that the bootstrap based $SADF_b$ test is much more computationally expensive in comparison to our proposed $STADF$. At the same time, the statistical conclusions based on the $STADF$ and $SADF_b$ tests do not contradict each other for all time series and for the significance level of 5\%, which indicates the obvious advantage of the $STADF$ test.

\section{Conclusion}
In this paper, we proposed the test for  bubbles under non-stationary volatility, which includes a finite number of jumps and a nonlinear shift of volatility. Our test is based on the sup-type $t$-statistics expanded under the null hypothesis, using the time-transformed data based on the variance profile. Due to this time-deformation technique, which was proposed  by \citet{CavaliereTaylor2007b}, we can treat the newly sampled data as if they were generated from the innovations with a constant variance. In fact, we showed that our test is asymptotically free of nuisance parameters and thus it can be conducted by using the asymptotic critical values. This implies that our test is computationally less expensive than the other existing bootstrap based tests in the literature. The finite sample performance showed that we can control the empirical size of our test relatively well and that our test performs better than the other existing tests robust to non-stationary volatility in view of power in some cases, although this is not always the case and it is less powerful in some other cases. Therefore, none of the existing tests outperforms the others uniformly. In this sense, our test can complement the existing ones and is useful to investigate the existence of a bubble under non-stationary volatility. In addition, our empirical example with the cryptocurrencies showed that the volatile behaviour of the time series can be partly explained by volatility changes and thus it is important to take volatility changes into account for such series.

\bibliographystyle{apalike} 
\bibliography{explosive,volatility}

\newpage

\newpage
\appendix
\renewcommand{\thesection}{\Alph{section}}
\renewcommand{\theequation}{\Alph{section}.\arabic{equation}}
\setcounter{section}{0}
\setcounter{equation}{0}

\section{Appendix: Proofs}

In this appendix, we denote C a generic constant that may differ from place to place. Also, because we subtract the initial value from all the observations, we set $y_0=0$ without loss of generality.

Let $T_{1,0}\coloneqq [\tau_{1,0}T]$, $T_{2,0}\coloneqq [\tau_{2,0}T]$ and $T_{3,0}\coloneqq [\tau_{3,0}T]$ and decompose the sample period such that $B_0\coloneqq [1, Th]$, $A_1\coloneqq [Th+1, T_{1,0}-Th]$, $B_1\coloneqq [T_{1,0}-Th+1,T_{1,0}+Th]$, $A_2\coloneqq [T_{1,0}+Th+1, T_{2,0}-Th]$, $B_2\coloneqq [T_{2,0}-Th+1,T_{2,0}+Th]$, $A_3\coloneqq [T_{2,0}+Th+1,T_{3,0}-Th]$, $B_3\coloneqq [T_{3,0}-Th+1,T_{3,0}+Th]$, $A_4\coloneqq [T_{3,0}+Th+1,T-Th]$, and $B_4\coloneqq [T-Th+1,T]$. We first state the asymptotic order of $\hat{\delta}_t$ given by \eqref{delta:lls}, which is derived in Lemma 1 of \citet{HLZ2021}. Note that \citet{HLZ2021} assumed the continuity of $\sigma_t$ throughout their paper, but this assumption is not used as far as the derivation of the following result is concerned and thus $\sigma_t$ can be discontinuous such as having a finite number of jumps.

\begin{lemma}
\label{lemma:phi}
Let $\phi_1=1+\delta_{1}$ and $\phi_2=1-\delta_{2}$. Under Assumptions \ref{assumption-1} and \ref{assumption:G},
\BEQA
\max_{Th+1\leq t \leq T_{1,0}-Th}|\hat{\delta}_t-\delta_t|
& = &
O_p\left(\sqrt{\frac{\log(T)}{T^2h}}\right),\\
\max_{T_{1,0}+Th+1\leq t \leq T_{2,0}-Th}|\phi_1^{t+Th-1-T_{1,0}}(\hat{\delta}_t-\delta_t)|
& = &
O_p\left(\frac{\log^2(T)}{T^{1/2-1/p}}\right),\\
\max_{T_{2,0}+Th+1\leq t \leq T_{3,0}-Th}|\phi_2^{t-Th-1-T_{2,0}}(\hat{\delta}_t-\delta_t)|
& = &
O_p\left(\frac{\log^2(T)}{\phi_1^{T_{2,0}-T_{1,0}}T^{1/2-1/p}}\right),\\
\max_{T_{3,0}+Th+1\leq t \leq T-Th}|\hat{\delta}_t-\delta_t|
& = &
O_p\left(\sqrt{\frac{\log(T)}{(a_T \vee \sqrt{T})^2Th}}\right).
\EEQA
\end{lemma}

The next lemma is concerned with the order of $y_t$, which is derived by using Lemma \ref{lemma:phi}.

\begin{lemma}
\label{lemma:y2}
Under Assumption \ref{assumption-1},
\[
y_t=\left\{\BA{lcl}
O_p(\sqrt{T}) & : & 1\leq t\leq T_{1,0}\\
\phi_1^{t-T_{1,0}}O_p(\sqrt{T}) & : & T_{1,0}+1 \leq t\leq T_{2,0}\\
\phi_2^{t-T_{2,0}}\phi_1^{T_{2,0}-T_{1,0}}O_p(\sqrt{T}) & : & T_{2,0}+1\leq t < T_{3,0}\\
O_p(a_T \vee \sqrt{T}) & : & T_{3,0}+1\leq t\leq T
\EA\right.
\]
where the orders are uniform over the corresponding range of $t$.
\end{lemma}
Proof: For $1\leq t\leq T_{1,0}$, we have
\[
y_t=y_0+\sum_{j=1}^t\varepsilon_t
\]
and thus, we obtain the result by \eqref{fclt1}.

For $T_{1,0}+1\leq t \leq T_{2,0}$, we can recursively solve $y_t$ as
\[
y_t=\phi_1^{t-T_{1,0}}\left(y_{T_{1,0}}+\sum_{j=T_{1,0}+1}^t\phi_1^{T_{1,0}-j}\varepsilon_j\right).
\]
Using Haj\'ek-R\'enyi inequality for martingale differences, we have
\BEQA
P\left(\max_{T_{1,0}+1\leq t\leq T_{2,0}}\left|\sum_{j=T_{1,0}+1}^t\phi_1^{T_{1,0}-j}\varepsilon_j\right|\geq M\right)
& \leq &
\frac{1}{M^2}\sum_{j=T_{1,0}+1}^{T_{2,0}}\phi_1^{2(T_{1,0}-j)}\sigma_j^2 \\
& \leq &
\frac{C}{M^2}\frac{1-\phi_1^{-2(T_{2,0}-T_{1,0})}}{\phi_1^2-1} < \frac{C}{M^2}
\EEQA
and thus $\max_{T_{1,0}+1\leq t\leq T_{2,0}}\left|\sum_{j=T_{1,0}+1}^t\phi_1^{T_{1,0}-j}\varepsilon_j\right|=O_p(1)$. Since $y_{T_{1,0}}=O_p(\sqrt{T})$, we have
\[
y_t=\phi_1^{t-T_{1,0}}(O_p(\sqrt{T})+O_p(1))
\]
uniformly. In particular, we have $y_{T_{2,0}}=\phi_1^{T_{2,0}-T_{1,0}}O_p(\sqrt{T})$.

Similarly, for $T_{2,0}+1\leq t < T_{3,0}$, we have
\[
y_t=\phi_2^{t-T_{2,0}}y_{T_{2,0}}+\sum_{j=T_{2,0}+1}^t\phi_2^{t-j}\varepsilon_j,
\]
where the second term on the right-hand side can be shown to be $O_p(1)$ uniformly because $|\phi_2| < 1$. Using the above result, we have $y_t=\phi_2^{t-T_{2,0}}\phi_1^{T_{2,0}-T_{1,0}}O_p(\sqrt{T})+O_p(1)$.

For $T_{3,0}\leq t \leq T$, $y_t=y_{T_{3,0}}+\sum_{j=T_{3,0}+1}^t\varepsilon_t$ and, using the assumption for $y_{T_{3,0}}$, we obtain the result.$\blacksquare$

\noindent
{\bf Proof of Theorem \ref{theorem-consistency}}: (i) is obtained from \eqref{df:num2}--\eqref{df:den2} and \eqref{fclt:2}.

(ii) To show the consistency, it is sufficient to show that $TADF_{r_1}^{r_2}$ given in \eqref{TT4} goes to infinity for a given specific value of $r_1$ and $r_2$. In particular, we consider the case where $ r_1 < \eta(\tau_{1,0}) < \eta(\tau_{2,0})= r_2$, which is equivalent to $g( r_1) < \tau_{1,0} < \tau_{2,0}=g( r_2)$.

For the numerator, we can see from Lemma \ref{lemma:y2} that
\begin{eqnarray}
\tilde{y}_{\lfloor  r_2T \rfloor}^2-\tilde{y}_{\lfloor  r_1T \rfloor}^2-\bar{\omega}^2(\lfloor  r_2T \rfloor-\lfloor  r_1T \rfloor)
& = &
{y}_{\lfloor \tau_{2,0}T \rfloor}^2-{y}_{\lfloor g(\lfloor  r_1T \rfloor/T)T\rfloor}^2-\bar{\omega}^2(\lfloor  r_2T \rfloor-\lfloor  r_1T \rfloor)
\nonumber \\
& = &
O_p\left(T\phi_1^{2(T_{2,0}-T_{1,0})}\right)-O_p(T)-O_p(T)
\nonumber \\
& = &
O_p\left(T\phi_1^{2(T_{2,0}-T_{1,0})}\right),
\label{consistency:numerator}
\end{eqnarray}
where the dominant term is positive.

On the other hand, we rewrite the summation in the numerator with respect to the original time ordering such that
\[
\sum_{t=\lfloor  r_1T \rfloor+1}^{\lfloor  r_2T \rfloor}\tilde{y}_{t-1}^2
=
\sum_{t=\lfloor  r_1T \rfloor+1}^{\lfloor  r_2T \rfloor}y_{\lfloor g((t-1)/T)T\rfloor}^2.
\]
Note that the summation is taken from $t=\lfloor  r_1T \rfloor+1$ to $\lfloor  r_2T \rfloor$ and thus the total number of summands is $\lfloor  r_2T\rfloor-\lfloor  r_1T\rfloor$, but, as we can see by focusing on the index given by $\lfloor g((t-1)/T)T\rfloor$, some of the original observations between $t=\lfloor g(\lfloor  r_1T \rfloor/T)T \rfloor$, $\lfloor g(\lfloor  r_1T \rfloor/T)T \rfloor+1,\ldots,\lfloor g((\lfloor  r_2T\rfloor-1)/T)T \rfloor$ may not appear as summands whereas the others may be summed up a few times. In particular, some of the non-time-transformed observations in the law volatility regime may be skipped but those in the high volatility regime may be added several times (but finite times). Taking this into account, we can see that for some constant $C$,
\begin{eqnarray*}
\sum_{t=\lfloor  r_1T \rfloor+1}^{\lfloor  r_2T \rfloor}\tilde{y}_{t-1}^2
& \leq &
C\sum_{t=\lfloor g(\lfloor  r_1T \rfloor/T)T\rfloor}^{\lfloor g(\lfloor  r_2T \rfloor/T)T\rfloor}y_{t}^2\\
& = &
C\sum_{t=\lfloor g(\lfloor  r_1T \rfloor/T)T\rfloor}^{\lfloor \tau_{1,0} T\rfloor}y_{t}^2+C\sum_{t=\lfloor \tau_{1,0}T\rfloor+1}^{\lfloor \tau_{2,0}T\rfloor}y_{t}^2\\
& = &
O_p(T^2)+O_p\left(T\phi_1^{2(T_{2,0}-T_{1,0})}\right)=O_p\left(T\phi_1^{2(T_{2,0}-T_{1,0})}\right),
\end{eqnarray*}
where the orders are obtained by using Lemma \ref{lemma:y2}. By combining this with \eqref{consistency:numerator}, we can finally have
\[
TADF_{r_1}^{r_2}=O_p\left(\sqrt{T}\phi_1^{T_{2,0}-T_{1,0}}\right)
\]
and the dominant term is positive.$\blacksquare$

\noindent

{\bf Proof of Proposition \ref{prop-consistency-estimator}}: (i) As explained in the proof of Theorem 2 of \citet{CavaliereTaylor2007a}, it is sufficient to prove the pointwise convergence. They showed that $\tilde{\eta}(s)\CP \eta(s)$ where $\tilde{\eta}(s)$ is defined as $\hat{\eta}(s)$ with $\hat{\varepsilon}_t^*$ replaced by $\varepsilon_t$. Then, the rest we have to show is the pointwise convergence given by
\BE
\left|\frac{1}{T}\sum_{t=1}^{\lfloor sT \rfloor}\hat{\varepsilon}_t^{*2}-\frac{1}{T}\sum_{t=1}^{\lfloor sT \rfloor}\varepsilon_t^2\right|\CP 0.
\label{conv:point}
\EE

Let ${\cal A}=A_1\cup A_2\cup A_3\cup A_4$ and ${\cal B}=B_0\cup B_1\cup B_2\cup B_3\cup B_4$. We consider \eqref{conv:point} on ${\cal A}$ and ${\cal B}$ separately. For brevity of notation, we write $t\in [1,\lfloor sT \rfloor]\cap {\cal A}$ as $t\in {\cal A}$ and $t\in [1,\lfloor sT \rfloor]\cap {\cal B}$ as $t\in {\cal B}$, respectively.

To show \eqref{conv:point} on ${\cal A}$, we first note that, because $\hat{\varepsilon}_t=-(\hat{\delta}_t-\delta_t)y_{t-1}+\varepsilon_t$,
\BEQA
\frac{1}{T}\sum_{t\in {\cal A}}\hat{\varepsilon}_t^{*2}
& = &
\frac{1}{T}\sum_{t\in {\cal A}}(\hat{\delta}_t-\delta_t)^2y_{t-1}^21(|\hat{\varepsilon}_t| < \psi_T) \\
& &
-\frac{2}{T}\sum_{t\in {\cal A}}(\hat{\delta}_t-\delta_t)y_{t-1}\varepsilon_t 1(|\hat{\varepsilon}_t| < \psi_T)
+
\frac{1}{T}\sum_{t\in {\cal A}}\varepsilon_t^2 1(|\hat{\varepsilon}_t| < \psi_T) \\
& = &
I+II+III.
\EEQA
We will show that $I\CP 0$ and $II\CP 0$, while $III-T^{-1}\sum \varepsilon_t^2\CP 0$.

To show $I\CP 0$, it is sufficient to show that
\BE
\left|I\right|\leq
\frac{1}{T}\sum_{t\in {\cal A}}(\hat{\delta}_t-\delta_t)^2y_{t-1}^2
\CP 0.
\label{conv:I}
\EE
Using Lemmas \ref{lemma:phi} and \ref{lemma:y2}, we have
\[
\max_{t\in A_1}\frac{1}{T}\sum_{i=Th+1}^t(\hat{\delta}_i-\delta_i)^2y_{i-1}^2
= O_p\left(\frac{\log (T)}{T^2h}\right)O_p(T)=O_p\left(\frac{\log (T)}{Th}\right)=o_p(1),
\]
\BEQA
\max_{t\in A_2}\frac{1}{T}\sum_{i=T_{1,0}+Th+1}^t(\hat{\delta}_i-\delta_i)^2y_{i-1}^2
& = &
\max_{t\in A_2}\frac{1}{T}\sum_{i=T_{1,0}+Th+1}^t\phi_1^{2(i+Th-1-T_{1,0})}(\hat{\delta}_i-\delta_i)^2\phi_1^{-2(i-T_{1,0})}y_{i-1}^2\phi_1^{-2(Th-1)} \\
& = &
O_p\left(\frac{\log^4 (T)}{T^{1-2/p}}\right)O_p(T)\phi_1^{-2(Th-1)} \\
& = &
O_p\left(\frac{T^{2/p}\log^4 (T)}{\phi_1^{2(Th-1)}}\right)=o_p(1),
\EEQA
\BEQA
\max_{t\in A_3}\frac{1}{T}\sum_{i=T_{2,0}+Th+1}^t(\hat{\delta}_i-\delta_i)^2y_{i-1}^2
& = &
\max_{t\in A_3}\frac{1}{T}\sum_{i=T_{2,0}+Th+1}^t\phi_2^{2(i-Th-1-T_{2,0})}(\hat{\delta}_i-\delta_i)^2\phi_2^{-2(i-T_{2,0})}y_{i-1}^2\phi_2^{2(Th+1)} \\
& = &
O_p\left(\frac{\log^4 (T)}{\phi_1^{2(T_{2,0}-T_{1,0})}T^{1-2/p}}\right)O_p\left(\phi_1^{2(T_{2,0}-T_{1,0})}T\right)\phi_2^{2(Th+1)} \\
& = &
O_p\left(\phi_2^{2(Th+1)}T^{2/p}\log^4 (T)\right)=o_p(1),
\EEQA
\[
\max_{t\in A_4}\frac{1}{T}\sum_{i=T_{3,0}+Th+1}^t(\hat{\delta}_i-\delta_i)^2y_{i-1}^2
= O_p\left(\frac{\log (T)}{(a_T \vee \sqrt{T})^2Th}\right)O_p\left((a_T \vee \sqrt{T})^2\right)=O_p\left(\frac{\log (T)}{Th}\right)=o_p(1).
\]
These results imply \eqref{conv:I}.

For $III$, we have
\BEQA
\frac{1}{T}\sum_{t\in {\cal A}}\varepsilon_t^2 1\left(|\hat{\varepsilon}_t| < \psi_T\right)-\frac{1}{T}\sum_{t\in {\cal A}}\varepsilon_t^2
& = &
\frac{1}{T}\sum_{t\in {\cal A}}\varepsilon_t^2 1\left(|\hat{\varepsilon}_t| \geq \psi_T\right) 
\nonumber \\
& \leq &
1\left(\max_{t\in {\cal A}} |\hat{\varepsilon}_t| \geq \psi_T\right)\frac{1}{T}\sum_{t\in {\cal A}}\varepsilon_t^2,
\EEQA
and thus, it is sufficient to show that 
\BE
1\left(\max_{t\in {\cal A}} |\hat{\varepsilon}_t| \geq \psi_T\right)=o_p(1).
\label{conv:IIIa}
\EE
Note that $\hat{\varepsilon}_t=-(\hat{\delta}_t-\delta_t)y_{t-1}+\varepsilon_t$. Similarly to the proof of the convergence of the term $I$, the first term on the right-hand side can be shown to be $o_p(1)$ uniformly over ${t\in {\cal A}}$ by using Lemmas \ref{lemma:phi} and \ref{lemma:y2}, while
\BEQA
P\left(\max_{t\in {\cal A}}|\varepsilon_t| \geq \psi_T \right)
& \leq &
\sum_{t\in {\cal A}}P\left(|\varepsilon_t|\geq \psi_T \right) \\
& \leq &
\frac{1}{\psi_T^p}\sum_{t\in {\cal A}}E|\varepsilon_t|^p \\
& \leq &
C\frac{T}{\psi_T^p}.
\EEQA
Because $T/\psi_T^p\to 0$ by Assumption \ref{assumption:psi}(b), \eqref{conv:IIIa} holds.

The convergence of $II$ is obtained by using \eqref{conv:I}, by noting that
\BEQA
|II|
& \leq &
\frac{2}{T}\sum_{t\in {\cal A}}\left|(\hat{\delta}_t-\delta_t)y_{t-1}\varepsilon_t\right|1\left(|\hat{\varepsilon}_t| < \psi_T\right) \\
& \leq &
2\sqrt{\frac{1}{T}\sum_{t\in {\cal A}}(\hat{\delta}_t-\delta_t)^2y_{t-1}^2}\sqrt{\frac{1}{T}\sum_{t\in {\cal A}}\varepsilon_t^2} \\
& = &
o_p(1).
\EEQA

To see \eqref{conv:point} on ${\cal B}$, we show that each term in \eqref{conv:point} converges to zero in probability on ${\cal B}$. Because the number of the observations on ${\cal B}$ is proportional to $Th$, we have
\[
\frac{1}{T}\sum_{t\in {\cal B}}\hat{\varepsilon}_t^{*2}=\frac{1}{T}\sum_{t\in {\cal B}}\hat{\varepsilon}_t^2 1\left(|\hat{\varepsilon}_t| < \psi_T\right) \\
\leq
C\frac{Th}{T}\psi_T^2=O(h\psi_T^2)=o(1),
\]
by Assumption \ref{assumption:psi}. Similarly, we can see that on $t\in {\cal B}$,
\[
\frac{1}{T}\sum_{t\in {\cal B}}\varepsilon_t^2=O_p(h)=o_p(1).
\]
We then obtain \eqref{conv:point} on ${\cal B}$.

(ii) As discussed in \citet{CavaliereTaylor2007b}, the result (i) implies the uniform convergence of $\hat{g}(s)$ to $g(s)$. Thus, following Theorem 1 of \citet{CavaliereTaylor2007b}, we have
\[
\frac{1}{\sqrt{T}}y_{\lfloor \hat{g}(s)T \rfloor}\Rightarrow \bar{\omega} W(s).
\blacksquare
\]

\newpage

\begin{table}[!h]
\centering
\footnotesize
\caption{Finite sample sizes and powers of nominal 0.05-level tests, single volatility shift, $\tau=0.5$\label{tab_1}}
\begin{tabularx}{1.03\textwidth}{cccccccccccccc} \toprule
$\sigma_1/\sigma_0$&$\delta_1$&\rotatebox{90}{$SADF$}&\rotatebox{90}{$SADF_b$}&\rotatebox{90}{$SBZ$}&\rotatebox{90}{$sSADF$}&\rotatebox{90}{$sSADF_u$}&\rotatebox{90}{$STADF$}&\rotatebox{90}{$SADF$}&\rotatebox{90}{$SADF_b$}&\rotatebox{90}{$SBZ$}&\rotatebox{90}{$sSADF$}&\rotatebox{90}{$sSADF_u$}&\rotatebox{90}{$STADF$}\\\hline
&&\multicolumn{6}{c}{$T=100$}&\multicolumn{6}{c}{$T=200$}\\
\cmidrule(r){3-8} \cmidrule(r){9-14}
$1/6$&0.00&0.012&0.027&0.065&0.043&0.033&0.059&0.029&0.047&0.081&0.037&0.050&0.040\\
&0.02&0.022&0.047&0.112&0.204&0.173&0.135&0.110&0.149&0.254&0.536&0.509&0.328\\
&0.04&0.099&0.164&0.254&0.375&0.348&0.301&0.587&0.615&0.638&0.753&0.747&0.664\\
&0.06&0.352&0.392&0.455&0.497&0.496&0.500&0.843&0.861&0.862&0.861&0.888&0.855\\
&0.08&0.566&0.608&0.628&0.556&0.637&0.642&0.933&0.935&0.936&0.923&0.950&0.929\\
&0.10&0.728&0.743&0.753&0.631&0.759&0.744&0.961&0.964&0.964&0.939&0.970&0.953\\\hline
$1/3$&0.00&0.015&0.039&0.065&0.036&0.038&0.036&0.025&0.036&0.062&0.039&0.030&0.048\\
&0.02&0.027&0.056&0.114&0.129&0.118&0.135&0.127&0.178&0.262&0.379&0.349&0.331\\
&0.04&0.112&0.166&0.275&0.301&0.293&0.327&0.581&0.610&0.628&0.668&0.677&0.652\\
&0.06&0.335&0.384&0.435&0.428&0.436&0.487&0.841&0.847&0.848&0.834&0.865&0.852\\
&0.08&0.597&0.632&0.653&0.559&0.649&0.662&0.921&0.924&0.926&0.885&0.932&0.925\\
&0.10&0.729&0.757&0.770&0.623&0.753&0.748&0.971&0.973&0.973&0.939&0.972&0.968\\\hline
1&0.00&0.022&0.032&0.064&0.039&0.038&0.047&0.033&0.039&0.067&0.042&0.043&0.049\\
&0.02&0.067&0.072&0.143&0.077&0.083&0.114&0.212&0.234&0.313&0.180&0.237&0.285\\
&0.04&0.202&0.219&0.312&0.181&0.228&0.264&0.605&0.615&0.639&0.476&0.604&0.595\\
&0.06&0.419&0.418&0.482&0.297&0.406&0.442&0.847&0.842&0.843&0.723&0.838&0.819\\
&0.08&0.615&0.630&0.663&0.464&0.608&0.625&0.936&0.939&0.940&0.839&0.938&0.926\\
&0.10&0.711&0.718&0.732&0.540&0.701&0.699&0.969&0.969&0.969&0.902&0.970&0.956\\\hline
3&0.00&0.284&0.057&0.077&0.040&0.072&0.051&0.366&0.063&0.090&0.056&0.076&0.063\\
&0.02&0.383&0.096&0.128&0.065&0.101&0.066&0.509&0.173&0.221&0.091&0.171&0.153\\
&0.04&0.470&0.182&0.228&0.101&0.165&0.161&0.718&0.441&0.485&0.296&0.440&0.406\\
&0.06&0.594&0.298&0.327&0.170&0.281&0.241&0.864&0.734&0.748&0.552&0.705&0.683\\
&0.08&0.699&0.459&0.485&0.295&0.429&0.365&0.940&0.873&0.876&0.754&0.860&0.845\\
&0.10&0.790&0.630&0.656&0.402&0.610&0.526&0.962&0.928&0.928&0.831&0.921&0.904\\\hline
6&0.00&0.540&0.084&0.101&0.049&0.079&0.045&0.608&0.070&0.094&0.037&0.067&0.061\\
&0.02&0.586&0.114&0.130&0.053&0.095&0.056&0.691&0.136&0.172&0.082&0.140&0.089\\
&0.04&0.632&0.171&0.194&0.090&0.168&0.080&0.792&0.337&0.374&0.244&0.343&0.219\\
&0.06&0.720&0.223&0.248&0.145&0.235&0.110&0.913&0.598&0.635&0.463&0.597&0.488\\
&0.08&0.770&0.350&0.375&0.232&0.334&0.174&0.952&0.777&0.796&0.650&0.759&0.703\\
&0.10&0.844&0.449&0.472&0.293&0.416&0.256&0.976&0.890&0.903&0.788&0.881&0.859\\\hline
  \bottomrule
\end{tabularx}
\end{table}

\begin{table}[!h]
\centering
\footnotesize
\caption{Finite sample sizes and powers of nominal 0.05-level tests, single volatility shift, $\tau=0.3$\label{tab_2}}
\begin{tabularx}{1.03\textwidth}{cccccccccccccc} \toprule
$\sigma_1/\sigma_0$&$\delta_1$&\rotatebox{90}{$SADF$}&\rotatebox{90}{$SADF_b$}&\rotatebox{90}{$SBZ$}&\rotatebox{90}{$sSADF$}&\rotatebox{90}{$sSADF_u$}&\rotatebox{90}{$STADF$}&\rotatebox{90}{$SADF$}&\rotatebox{90}{$SADF_b$}&\rotatebox{90}{$SBZ$}&\rotatebox{90}{$sSADF$}&\rotatebox{90}{$sSADF_u$}&\rotatebox{90}{$STADF$}\\\hline
&&\multicolumn{6}{c}{$T=100$}&\multicolumn{6}{c}{$T=200$}\\
\cmidrule(r){3-8} \cmidrule(r){9-14}
$1/6$&0.00&0.008&0.045&0.088&0.048&0.049&0.066&0.019&0.047&0.083&0.040&0.052&0.044\\
&0.02&0.006&0.039&0.086&0.302&0.249&0.157&0.027&0.101&0.245&0.634&0.587&0.395\\
&0.04&0.025&0.122&0.253&0.601&0.530&0.377&0.631&0.693&0.702&0.860&0.852&0.744\\
&0.06&0.324&0.433&0.480&0.725&0.679&0.560&0.843&0.868&0.870&0.934&0.926&0.863\\
&0.08&0.598&0.675&0.689&0.805&0.794&0.722&0.931&0.935&0.936&0.958&0.956&0.930\\
&0.10&0.735&0.794&0.797&0.850&0.854&0.789&0.965&0.968&0.968&0.968&0.978&0.959\\\hline
$1/3$&0.00&0.012&0.041&0.077&0.035&0.035&0.076&0.012&0.044&0.077&0.034&0.044&0.050\\
&0.02&0.004&0.029&0.098&0.157&0.123&0.158&0.056&0.136&0.254&0.403&0.358&0.350\\
&0.04&0.041&0.143&0.267&0.376&0.336&0.356&0.608&0.672&0.685&0.761&0.748&0.707\\
&0.06&0.338&0.439&0.483&0.576&0.546&0.549&0.836&0.860&0.864&0.887&0.884&0.858\\
&0.08&0.593&0.680&0.699&0.703&0.723&0.712&0.936&0.940&0.940&0.935&0.946&0.938\\
&0.10&0.729&0.772&0.776&0.760&0.794&0.765&0.969&0.970&0.970&0.950&0.970&0.963\\\hline
1&0.00&0.022&0.032&0.064&0.039&0.038&0.047&0.033&0.039&0.067&0.042&0.043&0.049\\
&0.02&0.067&0.072&0.143&0.077&0.083&0.114&0.212&0.234&0.313&0.180&0.237&0.285\\
&0.04&0.202&0.219&0.312&0.181&0.228&0.264&0.605&0.615&0.639&0.476&0.604&0.595\\
&0.06&0.419&0.418&0.482&0.297&0.406&0.442&0.847&0.842&0.843&0.723&0.838&0.819\\
&0.08&0.615&0.630&0.663&0.464&0.608&0.625&0.936&0.939&0.940&0.839&0.938&0.926\\
&0.10&0.711&0.718&0.732&0.540&0.701&0.699&0.969&0.969&0.969&0.902&0.970&0.956\\\hline
3&0.00&0.316&0.065&0.093&0.044&0.067&0.052&0.342&0.062&0.084&0.035&0.054&0.046\\
&0.02&0.378&0.098&0.141&0.057&0.105&0.116&0.523&0.157&0.245&0.106&0.163&0.215\\
&0.04&0.483&0.147&0.220&0.096&0.150&0.202&0.721&0.448&0.519&0.317&0.434&0.525\\
&0.06&0.599&0.294&0.378&0.193&0.294&0.353&0.859&0.753&0.772&0.584&0.746&0.768\\
&0.08&0.710&0.464&0.529&0.313&0.447&0.513&0.919&0.879&0.883&0.751&0.870&0.875\\
&0.10&0.797&0.626&0.679&0.408&0.593&0.651&0.966&0.945&0.947&0.848&0.943&0.945\\\hline
6&0.00&0.523&0.069&0.087&0.045&0.071&0.045&0.607&0.097&0.117&0.046&0.093&0.057\\
&0.02&0.556&0.094&0.133&0.054&0.083&0.100&0.702&0.125&0.237&0.108&0.125&0.223\\
&0.04&0.663&0.119&0.208&0.078&0.130&0.202&0.816&0.358&0.520&0.310&0.374&0.537\\
&0.06&0.708&0.193&0.331&0.169&0.215&0.334&0.906&0.668&0.752&0.537&0.657&0.765\\
&0.08&0.775&0.330&0.484&0.263&0.341&0.490&0.936&0.826&0.849&0.699&0.820&0.850\\
&0.10&0.847&0.509&0.621&0.357&0.496&0.616&0.974&0.934&0.940&0.829&0.934&0.938\\\hline
  \bottomrule
\end{tabularx}
\end{table}

\begin{table}[!h]
\centering
\footnotesize
\caption{Finite sample sizes and powers of nominal 0.05-level tests, single volatility shift, $\tau=0.7$\label{tab_3}}
\begin{tabularx}{1.03\textwidth}{cccccccccccccc} \toprule
$\sigma_1/\sigma_0$&$\delta_1$&\rotatebox{90}{$SADF$}&\rotatebox{90}{$SADF_b$}&\rotatebox{90}{$SBZ$}&\rotatebox{90}{$sSADF$}&\rotatebox{90}{$sSADF_u$}&\rotatebox{90}{$STADF$}&\rotatebox{90}{$SADF$}&\rotatebox{90}{$SADF_b$}&\rotatebox{90}{$SBZ$}&\rotatebox{90}{$sSADF$}&\rotatebox{90}{$sSADF_u$}&\rotatebox{90}{$STADF$}\\\hline
&&\multicolumn{6}{c}{$T=100$}&\multicolumn{6}{c}{$T=200$}\\
\cmidrule(r){3-8} \cmidrule(r){9-14}
$1/6$&0.000&0.026&0.036&0.071&0.045&0.045&0.058&0.037&0.047&0.080&0.042&0.048&0.048\\
&0.020&0.049&0.073&0.140&0.063&0.076&0.107&0.190&0.213&0.314&0.170&0.217&0.267\\
&0.040&0.167&0.211&0.308&0.139&0.191&0.257&0.585&0.606&0.628&0.469&0.588&0.591\\
&0.060&0.433&0.451&0.513&0.331&0.432&0.453&0.860&0.869&0.869&0.708&0.858&0.832\\
&0.080&0.580&0.609&0.637&0.428&0.587&0.584&0.940&0.944&0.944&0.855&0.938&0.924\\
&0.100&0.747&0.756&0.765&0.580&0.748&0.734&0.958&0.959&0.959&0.901&0.960&0.952\\\hline
$1/3$&0.000&0.024&0.037&0.075&0.030&0.043&0.054&0.035&0.042&0.088&0.045&0.051&0.055\\
&0.020&0.043&0.070&0.132&0.068&0.077&0.123&0.208&0.236&0.318&0.172&0.222&0.275\\
&0.040&0.158&0.186&0.281&0.139&0.197&0.241&0.617&0.642&0.663&0.497&0.628&0.623\\
&0.060&0.369&0.393&0.446&0.290&0.386&0.399&0.832&0.836&0.838&0.727&0.831&0.810\\
&0.080&0.595&0.610&0.634&0.448&0.588&0.598&0.926&0.931&0.931&0.846&0.928&0.921\\
&0.100&0.757&0.759&0.764&0.566&0.744&0.719&0.965&0.968&0.968&0.903&0.968&0.954\\\hline
1&0.000&0.022&0.032&0.064&0.039&0.038&0.047&0.033&0.039&0.067&0.042&0.043&0.049\\
&0.020&0.067&0.072&0.143&0.077&0.083&0.114&0.212&0.234&0.313&0.180&0.237&0.285\\
&0.040&0.202&0.219&0.312&0.181&0.228&0.264&0.605&0.615&0.639&0.476&0.604&0.595\\
&0.060&0.419&0.418&0.482&0.297&0.406&0.442&0.847&0.842&0.843&0.723&0.838&0.819\\
&0.080&0.615&0.630&0.663&0.464&0.608&0.625&0.936&0.939&0.940&0.839&0.938&0.926\\
&0.100&0.711&0.718&0.732&0.540&0.701&0.699&0.969&0.969&0.969&0.902&0.970&0.956\\\hline
3&0.000&0.256&0.072&0.092&0.037&0.071&0.053&0.294&0.075&0.098&0.030&0.066&0.060\\
&0.020&0.258&0.060&0.090&0.078&0.089&0.100&0.425&0.094&0.195&0.161&0.155&0.253\\
&0.040&0.403&0.131&0.204&0.168&0.187&0.238&0.721&0.486&0.541&0.495&0.523&0.610\\
&0.060&0.534&0.292&0.363&0.296&0.327&0.402&0.890&0.794&0.801&0.740&0.792&0.814\\
&0.080&0.704&0.522&0.548&0.430&0.512&0.563&0.954&0.923&0.924&0.868&0.917&0.922\\
&0.100&0.800&0.668&0.680&0.537&0.656&0.663&0.977&0.951&0.953&0.907&0.949&0.955\\\hline
6&0.000&0.496&0.080&0.099&0.039&0.075&0.070&0.560&0.064&0.088&0.033&0.059&0.053\\
&0.020&0.521&0.079&0.112&0.071&0.108&0.080&0.619&0.075&0.132&0.172&0.169&0.073\\
&0.040&0.534&0.077&0.126&0.154&0.156&0.073&0.828&0.367&0.408&0.496&0.496&0.250\\
&0.060&0.705&0.169&0.222&0.304&0.300&0.110&0.936&0.710&0.730&0.730&0.751&0.555\\
&0.080&0.788&0.389&0.424&0.437&0.459&0.219&0.975&0.887&0.897&0.860&0.889&0.782\\
&0.100&0.870&0.559&0.575&0.548&0.588&0.314&0.986&0.953&0.954&0.912&0.949&0.895\\\hline
\hline
  \bottomrule
\end{tabularx}
\end{table}

\begin{table}[!h]
\centering
\footnotesize
\caption{Finite sample sizes and powers of nominal 0.05-level tests, double volatility shift\label{tab_4}}
\begin{tabularx}{1.03\textwidth}{cccccccccccccc} \toprule
$\sigma_1/\sigma_0$&$\delta_1$&\rotatebox{90}{$SADF$}&\rotatebox{90}{$SADF_b$}&\rotatebox{90}{$SBZ$}&\rotatebox{90}{$sSADF$}&\rotatebox{90}{$sSADF_u$}&\rotatebox{90}{$STADF$}&\rotatebox{90}{$SADF$}&\rotatebox{90}{$SADF_b$}&\rotatebox{90}{$SBZ$}&\rotatebox{90}{$sSADF$}&\rotatebox{90}{$sSADF_u$}&\rotatebox{90}{$STADF$}\\\hline
&&\multicolumn{6}{c}{$T=100$}&\multicolumn{6}{c}{$T=200$}\\
\cmidrule(r){3-8} \cmidrule(r){9-14}
$1/6$&0.00&0.029&0.035&0.067&0.040&0.046&0.041&0.046&0.048&0.072&0.040&0.053&0.041\\
&0.02&0.040&0.052&0.127&0.364&0.332&0.178&0.122&0.160&0.287&0.651&0.643&0.377\\
&0.04&0.143&0.169&0.295&0.623&0.590&0.383&0.648&0.657&0.667&0.875&0.874&0.706\\
&0.06&0.376&0.419&0.474&0.746&0.719&0.524&0.848&0.852&0.855&0.937&0.934&0.844\\
&0.08&0.619&0.645&0.671&0.805&0.797&0.683&0.946&0.948&0.949&0.965&0.969&0.940\\
&0.10&0.762&0.771&0.776&0.831&0.852&0.769&0.974&0.974&0.974&0.982&0.984&0.969\\\hline
$1/3$&0.00&0.019&0.027&0.064&0.033&0.035&0.047&0.035&0.042&0.079&0.041&0.047&0.051\\
&0.02&0.030&0.042&0.137&0.196&0.182&0.162&0.147&0.172&0.303&0.469&0.457&0.362\\
&0.04&0.117&0.169&0.306&0.441&0.412&0.364&0.652&0.662&0.679&0.763&0.758&0.697\\
&0.06&0.399&0.430&0.486&0.606&0.595&0.527&0.863&0.866&0.871&0.896&0.903&0.865\\
&0.08&0.629&0.649&0.667&0.709&0.715&0.675&0.939&0.942&0.944&0.946&0.954&0.929\\
&0.10&0.762&0.789&0.795&0.783&0.815&0.775&0.975&0.977&0.977&0.961&0.976&0.969\\\hline
1&0.00&0.022&0.032&0.064&0.039&0.038&0.047&0.033&0.039&0.067&0.042&0.043&0.049\\
&0.02&0.067&0.072&0.143&0.077&0.083&0.114&0.212&0.234&0.313&0.180&0.237&0.285\\
&0.04&0.202&0.219&0.312&0.181&0.228&0.264&0.605&0.615&0.639&0.476&0.604&0.595\\
&0.06&0.419&0.418&0.482&0.297&0.406&0.442&0.847&0.842&0.843&0.723&0.838&0.819\\
&0.08&0.615&0.630&0.663&0.464&0.608&0.625&0.936&0.939&0.940&0.839&0.938&0.926\\
&0.10&0.711&0.718&0.732&0.540&0.701&0.699&0.969&0.969&0.969&0.902&0.970&0.956\\\hline
3&0.00&0.228&0.075&0.087&0.033&0.075&0.063&0.304&0.082&0.099&0.039&0.077&0.062\\
&0.02&0.318&0.117&0.150&0.053&0.110&0.108&0.460&0.200&0.257&0.090&0.187&0.193\\
&0.04&0.411&0.200&0.242&0.073&0.176&0.185&0.623&0.438&0.488&0.237&0.399&0.425\\
&0.06&0.525&0.301&0.364&0.144&0.270&0.323&0.808&0.711&0.734&0.507&0.688&0.710\\
&0.08&0.615&0.429&0.495&0.205&0.403&0.435&0.892&0.835&0.850&0.668&0.826&0.834\\
&0.10&0.713&0.562&0.608&0.305&0.526&0.582&0.944&0.923&0.926&0.785&0.920&0.912\\\hline
6&0.00&0.448&0.085&0.105&0.040&0.087&0.071&0.537&0.082&0.098&0.047&0.074&0.078\\
&0.02&0.528&0.126&0.172&0.051&0.121&0.123&0.663&0.149&0.224&0.070&0.132&0.176\\
&0.04&0.596&0.165&0.224&0.068&0.152&0.208&0.740&0.316&0.431&0.171&0.293&0.377\\
&0.06&0.636&0.223&0.299&0.097&0.200&0.274&0.836&0.567&0.650&0.401&0.545&0.634\\
&0.08&0.698&0.295&0.389&0.134&0.265&0.377&0.901&0.751&0.792&0.571&0.744&0.780\\
&0.10&0.761&0.448&0.527&0.224&0.404&0.487&0.955&0.891&0.905&0.684&0.885&0.907\\\hline
\hline
  \bottomrule
\end{tabularx}
\end{table}

\begin{table}[!h]
\centering
\footnotesize
\caption{Finite sample sizes and powers of nominal 0.05-level tests, logistic smooth transition in volatility\label{tab_5}}
\begin{tabularx}{1.03\textwidth}{cccccccccccccc} \toprule
$\sigma_1/\sigma_0$&$\delta_1$&\rotatebox{90}{$SADF$}&\rotatebox{90}{$SADF_b$}&\rotatebox{90}{$SBZ$}&\rotatebox{90}{$sSADF$}&\rotatebox{90}{$sSADF_u$}&\rotatebox{90}{$STADF$}&\rotatebox{90}{$SADF$}&\rotatebox{90}{$SADF_b$}&\rotatebox{90}{$SBZ$}&\rotatebox{90}{$sSADF$}&\rotatebox{90}{$sSADF_u$}&\rotatebox{90}{$STADF$}\\\hline
&&\multicolumn{6}{c}{$T=100$}&\multicolumn{6}{c}{$T=200$}\\
\cmidrule(r){3-8} \cmidrule(r){9-14}
$1/6$&0.00&0.021&0.047&0.083&0.043&0.056&0.057&0.026&0.031&0.064&0.046&0.044&0.049\\
&0.02&0.022&0.058&0.115&0.154&0.138&0.135&0.111&0.157&0.240&0.431&0.400&0.311\\
&0.04&0.092&0.150&0.254&0.335&0.307&0.323&0.605&0.637&0.665&0.714&0.720&0.690\\
&0.06&0.341&0.411&0.461&0.464&0.490&0.502&0.842&0.852&0.856&0.858&0.876&0.855\\
&0.08&0.591&0.642&0.663&0.578&0.653&0.670&0.913&0.922&0.922&0.898&0.933&0.912\\
&0.10&0.718&0.739&0.747&0.640&0.755&0.738&0.972&0.974&0.974&0.953&0.977&0.971\\\hline
$1/3$&0.00&0.014&0.034&0.067&0.028&0.040&0.050&0.019&0.038&0.083&0.051&0.051&0.059\\
&0.02&0.024&0.049&0.115&0.104&0.086&0.139&0.113&0.170&0.272&0.346&0.315&0.335\\
&0.04&0.114&0.158&0.266&0.263&0.249&0.320&0.605&0.639&0.663&0.678&0.683&0.685\\
&0.06&0.375&0.447&0.492&0.411&0.470&0.517&0.852&0.865&0.866&0.824&0.875&0.863\\
&0.08&0.559&0.605&0.635&0.538&0.628&0.639&0.942&0.946&0.946&0.913&0.954&0.944\\
&0.10&0.728&0.754&0.760&0.611&0.748&0.748&0.967&0.972&0.972&0.924&0.972&0.964\\\hline
1&0.00&0.022&0.032&0.064&0.039&0.038&0.047&0.033&0.039&0.067&0.042&0.043&0.049\\
&0.02&0.067&0.072&0.143&0.077&0.083&0.114&0.212&0.234&0.313&0.180&0.237&0.285\\
&0.04&0.202&0.219&0.312&0.181&0.228&0.264&0.605&0.615&0.639&0.476&0.604&0.595\\
&0.06&0.419&0.418&0.482&0.297&0.406&0.442&0.847&0.842&0.843&0.723&0.838&0.819\\
&0.08&0.615&0.630&0.663&0.464&0.608&0.625&0.936&0.939&0.940&0.839&0.938&0.926\\
&0.10&0.711&0.718&0.732&0.540&0.701&0.699&0.969&0.969&0.969&0.902&0.970&0.956\\\hline
3&0.00&0.308&0.066&0.089&0.042&0.079&0.050&0.342&0.075&0.095&0.034&0.065&0.051\\
&0.02&0.379&0.093&0.118&0.052&0.087&0.080&0.496&0.170&0.216&0.100&0.164&0.147\\
&0.04&0.465&0.185&0.226&0.097&0.172&0.146&0.723&0.480&0.523&0.268&0.448&0.429\\
&0.06&0.578&0.304&0.354&0.178&0.295&0.240&0.870&0.749&0.761&0.557&0.731&0.696\\
&0.08&0.700&0.479&0.510&0.264&0.438&0.387&0.920&0.855&0.861&0.691&0.842&0.809\\
&0.10&0.789&0.630&0.643&0.382&0.600&0.540&0.977&0.954&0.954&0.846&0.944&0.923\\\hline
6&0.00&0.512&0.073&0.095&0.036&0.072&0.054&0.587&0.082&0.102&0.033&0.070&0.049\\
&0.02&0.601&0.119&0.144&0.052&0.112&0.074&0.712&0.160&0.198&0.069&0.149&0.110\\
&0.04&0.641&0.172&0.198&0.075&0.153&0.086&0.792&0.306&0.349&0.175&0.286&0.200\\
&0.06&0.721&0.247&0.281&0.124&0.227&0.125&0.896&0.605&0.634&0.386&0.578&0.507\\
&0.08&0.782&0.325&0.364&0.170&0.299&0.168&0.949&0.797&0.811&0.611&0.774&0.731\\
&0.10&0.840&0.440&0.477&0.270&0.426&0.275&0.973&0.905&0.910&0.766&0.894&0.869\\\hline
  \bottomrule
\end{tabularx}
\end{table}

\begin{table}[!h]
\centering
\footnotesize
\caption{Finite sample sizes and powers of nominal 0.05-level tests, trending volatility\label{tab_6}}
\begin{tabularx}{1.03\textwidth}{cccccccccccccc} \toprule
$\sigma_1/\sigma_0$&$\delta_1$&\rotatebox{90}{$SADF$}&\rotatebox{90}{$SADF_b$}&\rotatebox{90}{$SBZ$}&\rotatebox{90}{$sSADF$}&\rotatebox{90}{$sSADF_u$}&\rotatebox{90}{$STADF$}&\rotatebox{90}{$SADF$}&\rotatebox{90}{$SADF_b$}&\rotatebox{90}{$SBZ$}&\rotatebox{90}{$sSADF$}&\rotatebox{90}{$sSADF_u$}&\rotatebox{90}{$STADF$}\\\hline
&&\multicolumn{6}{c}{$T=100$}&\multicolumn{6}{c}{$T=200$}\\
\cmidrule(r){3-8} \cmidrule(r){9-14}
$1/6$&0.00&0.002&0.032&0.061&0.027&0.037&0.053&0.005&0.037&0.075&0.035&0.041&0.054\\
&0.02&0.012&0.054&0.139&0.097&0.086&0.119&0.087&0.196&0.306&0.229&0.250&0.301\\
&0.04&0.093&0.197&0.333&0.226&0.238&0.309&0.605&0.658&0.672&0.608&0.660&0.659\\
&0.06&0.327&0.437&0.511&0.376&0.439&0.480&0.842&0.858&0.861&0.793&0.857&0.844\\
&0.08&0.602&0.677&0.692&0.558&0.661&0.645&0.929&0.938&0.938&0.888&0.936&0.926\\
&0.10&0.731&0.782&0.792&0.634&0.777&0.757&0.963&0.970&0.970&0.910&0.967&0.952\\\hline
$1/3$&0.00&0.009&0.029&0.066&0.026&0.028&0.058&0.008&0.038&0.076&0.035&0.043&0.043\\
&0.02&0.011&0.056&0.112&0.070&0.068&0.111&0.111&0.201&0.293&0.223&0.234&0.278\\
&0.04&0.119&0.208&0.323&0.202&0.236&0.291&0.579&0.640&0.656&0.568&0.631&0.630\\
&0.06&0.383&0.460&0.515&0.388&0.466&0.497&0.822&0.844&0.847&0.773&0.839&0.828\\
&0.08&0.579&0.630&0.651&0.503&0.621&0.615&0.931&0.940&0.942&0.883&0.941&0.927\\
&0.10&0.721&0.763&0.770&0.619&0.760&0.729&0.961&0.965&0.965&0.906&0.963&0.956\\\hline
1&0.00&0.022&0.032&0.064&0.039&0.038&0.047&0.033&0.039&0.067&0.042&0.043&0.049\\
&0.02&0.067&0.072&0.143&0.077&0.083&0.114&0.212&0.234&0.313&0.180&0.237&0.285\\
&0.04&0.202&0.219&0.312&0.181&0.228&0.264&0.605&0.615&0.639&0.476&0.604&0.595\\
&0.06&0.419&0.418&0.482&0.297&0.406&0.442&0.847&0.842&0.843&0.723&0.838&0.819\\
&0.08&0.615&0.630&0.663&0.464&0.608&0.625&0.936&0.939&0.940&0.839&0.938&0.926\\
&0.10&0.711&0.718&0.732&0.540&0.701&0.699&0.969&0.969&0.969&0.902&0.970&0.956\\\hline
3&0.00&0.178&0.046&0.071&0.037&0.050&0.042&0.186&0.050&0.076&0.044&0.059&0.054\\
&0.02&0.227&0.087&0.133&0.063&0.084&0.100&0.405&0.187&0.270&0.127&0.189&0.239\\
&0.04&0.358&0.161&0.237&0.122&0.161&0.202&0.687&0.564&0.599&0.380&0.551&0.555\\
&0.06&0.511&0.357&0.413&0.219&0.341&0.379&0.857&0.801&0.811&0.640&0.787&0.780\\
&0.08&0.657&0.535&0.571&0.358&0.517&0.525&0.924&0.898&0.901&0.782&0.894&0.884\\
&0.10&0.738&0.650&0.674&0.455&0.636&0.625&0.958&0.947&0.949&0.862&0.945&0.933\\\hline
6&0.00&0.291&0.060&0.071&0.038&0.056&0.041&0.340&0.053&0.076&0.032&0.050&0.042\\
&0.02&0.340&0.077&0.127&0.065&0.079&0.098&0.510&0.175&0.250&0.124&0.169&0.211\\
&0.04&0.498&0.175&0.241&0.110&0.149&0.186&0.722&0.499&0.542&0.345&0.471&0.502\\
&0.06&0.551&0.308&0.360&0.175&0.295&0.305&0.874&0.765&0.780&0.578&0.748&0.744\\
&0.08&0.697&0.492&0.530&0.307&0.461&0.494&0.949&0.901&0.905&0.762&0.895&0.880\\
&0.10&0.812&0.635&0.668&0.430&0.603&0.610&0.980&0.962&0.963&0.858&0.961&0.950\\\hline
  \bottomrule
\end{tabularx}
\end{table}

\begin{table}[!h]
\centering
\footnotesize
\caption{Empirical results, $p$-values\label{tab_emp}}
\begin{tabularx}{0.53\textwidth}{cccccccccc} \toprule
Series&$SADF$&$SADF_b$&$SBZ$&$sSADF$&$STADF$\\\hline
btc&0.000&0.083&0.109&0.093&0.099\\
eth&0.004&0.081&0.125&0.190&0.486\\
xrp&0.951&0.936&0.935&0.125&0.915\\
xlm&0.837&0.753&0.643&0.379&0.335\\
bch&0.000&0.072&0.109&0.146&0.189\\
ltc&0.042&0.228&0.252&0.439&0.186\\
eos&0.093&0.284&0.277&0.656&0.191\\
bnb&0.013&0.209&0.070&0.414&0.108\\
ada&0.000&0.046&0.065&0.086&0.044\\
xtz&0.000&0.059&0.099&0.109&0.373\\
etc&0.027&0.139&0.234&0.350&0.373\\
xmr&0.119&0.361&0.398&0.501&0.442\\
\hline
  \bottomrule
\end{tabularx}
\end{table}

\begin{table}[!h]
\centering
\footnotesize
\caption{Empirical results for large sample sizes, $p$-values and computational time (in seconds)\label{tab_emp2}}
\begin{tabular}{cccccccccc}
\hline \toprule
\multirow{2}{*}{Series} & \multirow{2}{*}{Start date} & \multirow{2}{*}{End date} & \multirow{2}{*}{$T$} & \multicolumn{2}{c}{$SADF$} & \multicolumn{2}{c}{$SADF_b$} & \multicolumn{2}{c}{$STADF$} \\ \cline{5-10} 
                        &                             &                           &                      & time      & $p$-value      & time        & $p$-value      & time       & $p$-value      \\ \hline
btc                     & 2010-07-18                  & 2021-06-01                & 3972                 & 1.59      & 0.000          & 336.31      & 0.000          & 6.21       & 0.044          \\
eth                     & 2015-08-08                  & 2021-06-01                & 2125                 & 0.48      & 0.000          & 104.60      & 0.000          & 2.09       & 0.011          \\
xrp                     & 2014-08-15                  & 2021-06-01                & 2483                 & 0.63      & 0.000          & 133.60      & 0.000          & 2.78       & 0.000          \\
xlm                     & 2015-09-30                  & 2021-06-01                & 2072                 & 0.44      & 0.000          & 91.32       & 0.000          & 2.03       & 0.000          \\
ltc                     & 2013-04-01                  & 2021-06-01                & 2984                 & 1.00      & 0.000          & 213.59      & 0.000          & 3.81       & 0.000          \\ \hline
\bottomrule
\end{tabular}
\end{table}

\end{document}